\documentclass[utf8]{frontiersSCNS} 

\usepackage{url,hyperref,lineno,microtype,subcaption}
\usepackage[onehalfspacing]{setspace}
\usepackage{natbib}
\usepackage[T1]{fontenc}
\usepackage{aas_macros}

\usepackage{ae,aecompl}
\usepackage{graphicx}   
\usepackage{amsmath}    
\usepackage{amssymb}

\def\sigs{\mbox{$\sigma_\star$}}
\def\sige{\mbox{$\sigma_{\rm e}$}}

\def\rhoDM{\mbox{$\langle \rho_{\rm DM} \rangle$}}

\def\Msun{\mbox{$M_\odot$}}

\def\ML{\mbox{$M/L$}}

\def\Mdyn{\mbox{$M_{\rm dyn}$}}

\def\mst{\mbox{$M_{\star}$}}

\def\Mvir{\mbox{$M_{\rm vir}$}}
\def\cvir{\mbox{$c_{\rm vir}$}}

\def\fdm{\mbox{$f_{\rm DM}$}}

\def\lsim{\mathrel{\rlap{\lower3.5pt\hbox{\hskip0.5pt$\sim$}}
    \raise0.5pt\hbox{$<$}}}                
\def\gsim{~\rlap{$>$}{\lower 1.0ex\hbox{$\sim$}}}
\def\Rap{\mbox{$R_{\rm Ap}$}}
\def\sigAp{\mbox{$\sigma_{\rm Ap}$}}
\def\TtoSM{\mbox{$M_{\rm dyn}/\mst$}}

\def\Fig{\mbox{Figure~}}
\def\Figs{\mbox{Figures~}}

\def\Sec{\mbox{Section~}}

\def\Re{\mbox{$R_{\rm e}$}}

\def\sqd{\mbox{~sq.~deg.}}

\def\keyFont{\fontsize{8}{11}\helveticabold }
\def\firstAuthorLast{Tortora \& Napolitano} 
\def\Authors{C.~Tortora\,$^{2}$ and N.~R.~Napolitano\,$^{1,2,*}$}
\def\Address{$^{1}$School of Physics and Astronomy, Sun Yat-sen University
Zhuhai Campus, Daxue Road 2, 519082 - Tangjia, Zhuhai, Guangdong,
P.R. China\\
$^{2}$INAF -- Osservatorio Astronomico di Capodimonte, Salita Moiariello 16, 80131 - Napoli, Italy }
\def\corrAuthor{N.R. Napolitano}

\def\corrEmail{napolitano@mail.sysu.edu.cn}

\begin{document}
\onecolumn
\firstpage{1}

\title[Dark matter in early-type galaxies]{The central dark matter fraction of massive early-type galaxies}

\author[\firstAuthorLast ]{\Authors} 
\address{} 
\correspondance{} 

\extraAuth{}

\maketitle

\begin{abstract}


Dark matter (DM) is predicted to be the dominant mass component in galaxies. In the central region of Early-type galaxies it is expected to account for a large amount of the total mass, although the stellar mass should still represent the majority of the mass budget,
depending on the stellar Initial Mass Function (IMF).
We discuss latest results on the DM fraction and mean DM density for local galaxies and explore their evolution with redshifts in the last 8 Gyr of the cosmic history. We compare these results with expectations from the $\Lambda$CDM model, and discuss the the role of the IMF and galaxy model, through the central total mass density slope. We finally present future perspectives offered by next generation instruments/surveys (Rubin/LSST, Euclid, CSST, WEAVE, 4MOST, DESI), that will provide the unique chance to measure the DM evolution with time for an unprecedented number of galaxies and constrain their evolutionary scenario.

\tiny
 \keyFont{ \section{Keywords:} dark matter; galaxies: formation; galaxy: evolution} 
\end{abstract}

\section{Introduction}

Early-type galaxies (ETGs) are the most massive galaxy systems in the Universe and represent the final stage of galaxy evolution. As such, they
carry the fossil record of the assembly of stellar and dark matter through
time, and, because they are bright and massive, can be
studied in details from low to high redshifts. They are the result of a complex alchemy of  physical processes that can be grouped in two major categories: galaxy merging (\citealt{Trujillo+06,Trujillo+07}; \citealt{Cenarro_Trujillo09}; \citealt{Naab+09}; \citealt{Hopkins+10_Mergers_LCDM}) and feedback mechanisms (e.g. \citealt{Dekel_Burkert14}; \citealt{dek_birn06}). The variety of evolutionary mechanisms they have experienced make them a crucial test bench of the galaxy formation theories and the ultimate probe of the hierarchical evolutionary scenario. In particular, they can be used to trace the assembly of both the luminous and the dark components across time and check if the observations are consistent with predictions from simulations.
For example, the total stellar-to-dark mass ratio of ETGs depends strongly on the galaxy mass, and observational and theoretical studies have suggested this to be connected to the overall star formation efficiency (\citealt{Benson+00}; \citealt{MH02};
\citealt{Napolitano+05}; \citealt{Mandelbaum+06};
\citealt{vdB+07}; \citealt{CW09}; \citealt{Moster+10};
\citealt{Alabi+16}).

The ETG centers should experience, on small scales, the effect of the mechanisms acting at the large scales and show similar correlations for the central masses as the ones involving their total masses. In the last two decades,
increasing evidences have been collected, showing that the central DM fraction (typically within one effective radius, \Re\ hereafter) is higher in larger and more massive galaxies (e.g. \citealt{HB09_FP}; \citealt{Tortora+09}; \citealt{RS09};
\citealt{Auger+10_SLACSX}; \citealt{NRT10}; \citealt{ThomasJ+11};
\citealt{SPIDER-VI}).

This positive correlation with mass seems almost insensitive to the
adopted galaxy mass profile or initial mass function, IMF (e.g.,
\citealt{Cardone+09}; \citealt{CT10}; \citealt{Cardone+11SIM}),
but it can become uncertain in case a non-$\Lambda$CDM scenario,
with mass following the (non-homologous) light distribution, is
adopted (e.g., \citealt{TBB04}; \citealt{Tortora+09,SPIDER-VI}). In general, a general consensus about the existence of such a trend is not yet reached as there are sparse evidences of an anti-correlation of the dark matter fraction with mass  (e.g., \citealt{Grillo+09, Grillo10,
Grillo_Cobat10}).

Among the largest source of uncertainty to quantify the stellar and DM mass budget in the central galactic regions, the IMF remains the most difficult to account for. Direct constraints from gravity sensitive spectral lines (see \citealt{Spiniello+12}; \citealt{LaBarbera+13_SPIDERVIII_IMF}) are strongly required, albeit quite expensive in terms of observational requirements.
In absence of these constraints, the simplest approach is to assume a ``universal'' IMF. Unfortunately, this can introduce unreasonable fluctuations of the stellar mass involved in the models, i.e. by a factor of 2 or more (e.g. assuming a
\citealt{Chabrier01} or a \citealt{Salpeter55} IMF or even
super-Salpeter IMF, e.g. \citealt{Tortora+09}). This can strongly
affect the conclusions on the central DM fraction in these
extreme cases. On the other hand, ``non-universality'' of the IMF, i.e. the systematic variation of the IMF with mass (or velocity dispersion), from a bottom-lighter (i.e., 'lower-mass') IMF for low mass ETGs to a bottom-heavier (i.e., 'higher-mass') IMF in massive galaxies, has accumulated convincing evidences in the last decade (e.g., \citealt{Treu+10};
\citealt{Conroy_vanDokkum12b}; \citealt{Cappellari+12},
\citealt{Spiniello+12};
\citealt{Goudfrooij_Kruijssen13};
\citealt{LaBarbera+13_SPIDERVIII_IMF};
\citealt{TRN13_SPIDER_IMF};
\citealt{Martin-Navarro+15_IMF_variation};
\citealt{Li+17_IMF}; \citealt{2019MNRAS.489.5612D}), despite, also in this case, contradicting evidences pointing to a bottom-light IMFs for massive systems have been reported (\citealt{Smith+15_SINFONI}). This non universal IMF can, in principle,
incorporate most (if not all) the ``apparent'' DM fraction trend with mass  (e.g., \citealt{ThomasJ+11}; \citealt{TRN13_SPIDER_IMF}).

A still almost unexplored area, to test the correlation of the dark-matter fraction with mass, is represented by high-redshift (i.e. $z>0.3$) galaxies. For these systems, the difficulty of obtaining extended kinematics over large samples of galaxies is a barrier for large statistical studies. Here, one possibility is the use of strong gravitational lensing (\citealt{Auger+09_SLACSIX, Auger+10_SLACSX},
\citealt{Tortora+10lensing}; \citealt{Sonnenfeld+13_SL2S_IV}). However, at the moment the number of available systems with all the necessary data to perform accurate mass models is still restricted to small samples and small redshift windows, i.e. $z\lsim0.7$ (e.g. \citealt{Shajib+21}).
Another option is to use simple dynamical analysis based on integrated aperture kinematics, as the one provided by large sky surveys, like SDSS/BOSS (e.g. \citealt{Thomas+13_BOSS}), GAMA (e.g. \citealt{Liske+15_GAMA}) and LAMOST (\citealt{Napolitano+20_LAMOST}). It has been proven that in these cases Jeans equation analysis can produce results consistent with more sophisticated dynamical modeling (\citealt{Tortora+09}, see also \Sec\ref{sec:general}).

The first systematic studies of the evolution of the central DM
fraction with redshift has been performed by
\cite{Beifiori+14} and \cite{Tortora+14_DMevol}, which provided
evidences that high$-z$ ETGs are less DM dominated than their
local counterparts. Only recently, we have updated our previous analysis
to include the effect of the IMF in the dynamical analysis and discussed in more details the results in the context of the hierarchical scenario (\citealt{Tortora+18_KiDS_DMevol}).

The paucity of high$-z$ studies remains a strong limitation in the understanding of the DM content in galaxies. In fact, by studying correlations at higher-redshift we could possibly separate the effect of the DM assembly and the IMF non-universality more clearly, when one or either effects start to emerge or the freedom on the choice of some parameters (e.g. age, metallicity of stars, concentration of the DM haloes, etc.) is lower.

In this paper we present some updated results about the DM content of the central regions of massive ETGs, compare them with other observations and simulations, and discuss some perspectives in the context of the upcoming sky surveys.
Despite in the last decade there has been a significant production of studies addressing these specific measurements in ETGs (\citealt{Padmanabhan+04}; \citealt{Cappellari+06,Cappellari+13_ATLAS3D_XV}; \citealt{HB09_FP}; \citealt{Tortora+09,Tortora+10CG,SPIDER-VI,Tortora+14_DMevol,Tortora+14_DMslope}; \citealt{ThomasJ+09,ThomasJ+11}; \citealt{RS09};
\citealt{Auger+10_SLACSX}; \citealt{Cardone+09,CT10}; \citealt{NRT10}; \citealt{Beifiori+14}; \citealt{Shu+15_SLACSXII}; \citealt{Nigoche-Netro+16,Nigoche-Netro+19}; \citealt{Xu+17_Illustris}; \citealt{Lovell+18_Illustris}), only few reviews on dark matter in galaxies have touched this specific subject (e.g. \citealt{Courteau+14_review}; \citealt{Salucci+19_DM}), with little emphasis on the insight the dark matter fractions can provide about the ETG structure and evolution.

The paper outline is as follows. In \Sec\ref{sec:general} we
present the procedure we have used to determine the DM content and
datasamples. We correlate the DM fraction with galaxy parameters
in \Sec\ref{sec:DM_correlations}, deriving hints on the DM and
total mass density slopes in \Sec\ref{sec:av_DM}. The evolution of
DM fraction with redshift and the interpretation within the galaxy
evolution scenario is provided in \Sec\ref{sec:DM_evolution}. Alternatives to the $\Lambda$CDM are briefly touched on in \Sec\ref{sec:alternatives}. A
summary of the results and future prospects are presented in
\Sec\ref{sec:conclusions}.

\section{Estimating the DM content}\label{sec:general}

We start by defining the dynamical mass within a given galaxy radius, $r$. This is historically
estimated using a simplistic ``virial formula''
\begin{equation}
M_{dyn}= \frac{K \sigma^{2} r}{G},\label{eq:M_virial}
\end{equation}
where  $G$  is the  gravitational  constant,  $\sigma$  is the
galaxy velocity dispersion within some aperture (e.g. circular from fiber spectroscopy, or rectangular from multi-slit spectroscopy), and  $K$  is a, so called,  pressure correction term  (e.g. \citealt{Padmanabhan+04}; \citealt{Cappellari+06};
\citealt{Tortora+09,SPIDER-VI}). This correction term depends on several
factors, including the radius where the dynamical mass, \Mdyn \, is computed in, the
aperture used to measure the $\sigma$,  the viewing angle of the
system, the orbital distribution of stars in the same radius, the luminosity profile slope,  and, finally, the DM density profile properties.

This explains why this approach, albeit simple and quite diffuse,
does not really capture the full intrinsic dynamics of a given galaxy, and usually some approximated expressions are used to account for average behaviours of the systems under exam.\\

\subsection{Jeans equations and total mass determination}\label{subsec:Jeans}
More precise mass estimates come from the direct modeling of individual galaxies using the Jeans equations. In this case one can fully take into account the spatial distribution of the stars, e.g. from some detailed modeling of their surface brightness (e.g. using a S\'ersic profile, \citealt{Sersic68}), assume some anisotropy of the stellar orbits, and finally assume different DM density profiles, hence testing different DM models.  With this approach one can finally estimate the model parameters of the dynamical mass, \Mdyn , within any radius (although the usual choice is to take $r=$~1~\Re) by solving the Jeans equation. In particular, one can derive the mass model parameters by best-fitting the velocity dispersion, projected along the line-of-sight and integrated within a given aperture, to the observed one from fiber or long-slit spectroscopy via $\chi^2$ minimization (see e.g. \citealt{Tortora+09,CT10,Tortora+10lensing,SPIDER-VI}). In details:
\begin{enumerate}
\item The projected luminosity profile $I(r)$ is parameterized by a
\citet{Sersic68} model.
\item The total cumulative (deprojected) dynamical mass
profile $M(r)$ is assumed to follow different distributions: a) a constant-$M/L$ profile $M(r) =\Upsilon_0 \, L(r)$, where $L(r)$ is the cumulative luminosity profile, e.g. from the curve of growth of the surface brightness profile $I(r)$, and $\Upsilon_0$ is the constant stellar mass-to-light ratio, or b) a singular isothermal sphere (SIS), where
$M(r)\propto \sigma_{\rm SIS}^{2} r$. In both cases the parameterized profiles have a single free parameter ($\Upsilon_0$ or $\sigma_{\rm SIS}$, respectively). The use of SIS model, in particular, is
motivated by evidences from strong gravitational lensing, showing that the total mass density of massive ETGs is close to a power law with slope equal to $-2$  (e.g.
\citealt{Koopmans+06_SLACSIII, Koopmans+09}.
\citealt{Gavazzi+07_SLACSIV}). More complex two-component models
can also be adopted, e.g. using DM profiles from N-body
simulations, as the NFW double power-law (\citealt{NFW96}), its
adiabatically contracted version (e.g. \citealt{Gnedin+04}), or
cored DM models (e.g. \citealt{Burkert95}). Exploring different
kinds of (more general) models (e.g. \citealt{Tortora+07}; \citealt{Cardone+09, Cardone+11SIM}) can finally help probing how the total mass distribution slope can change in terms of mass and galaxy
parameters (e.g. \citealt{Tortora+14_DMslope}).
\item The Jeans equation:
\begin{equation}
{{\rm d}(j_* \sigma_r^2) \over {\rm d}r} + 2\,{\beta(r) \over r}
\,j_* \sigma_r^2 = - j_*(r)\, \frac{GM(r)}{r^2} \ ,
\label{eq:jeans}
\end{equation}
where $\beta = 1 - \sigma_t^2/\sigma_r^2$ is the anisotropy parameter, is solved. Implicitly, the adoption of the spherical Jeans equation
assumes spherical symmetry and no rotation (cf.
\citealt{ML05a}). When isotropy ($\beta=0$) is adopted, the Jeans
Eq.~(\ref{eq:jeans}) simply writes as:
\begin{equation}
\sigma_r^2 (r) = \frac{1}{j_*(r)} \int_r^\infty j_* \frac{GM}{s^2}
{\rm d}s \ . \label{eq:iso}
\end{equation}
\item Eq. (\ref{eq:iso}) is projected along the line-of-sight to obtain the projected velocity dispersion:
\begin{equation}
\sigma_{\rm los}^2 (R) = \frac{2}{I(R)}\,\int_R^\infty \frac{j_*
\sigma_r^2 \,r\,{\rm d}r}{\sqrt{r^2\!-\!R^2}} , \label{eq:siglos}
\end{equation}
where
\begin{equation}
I(R) = 2\,\int_R^\infty \frac{j_*\,r}{\sqrt{r^2\!-\!R^2}}{\rm d}r
\label{eq:IR}
\end{equation}
is the projected density profile.
\item $\sigma_{\rm los}$ is integrated within a fixed aperture \Rap\ to obtain the aperture velocity dispersion, \sigAp\
using the Equation:
\begin{equation}
\sigma_{\rm Ap}^2 (\Rap) = \frac{1}{L(\Rap)}\int_0^{\Rap} 2\pi\,s\,I(s)\,\sigma_{\rm los}^2(s)\,{\rm d}s \
, \label{eq:sigap}
\end{equation}
where $L(R) = \int_0^R 2\pi s I(s)\, {\rm d}s$ is the luminosity
within the projected radius $R$, and $s$ is the generic coordinate along the los. To avoid lengthy calculations, we
have adopted the compact formulae for \sigAp\ calculated in
Equation B7 of \cite{ML05a} (note that the correct version of Equation B7 is reported in \citealt{ML06_erratum}).
\item The modelled $\sigma_{\rm Ap}$ derived above is fitted to the observed velocity dispersion, $\sigma$, by varying the mass model free parameters until the best-fit is achieved. In this paper, we will adopt one-parameter mass models. As matter of fact, even in the case of multi-parametric models, we will reduce these to a single parameter, by using realistic correlations among the model parameters. In particular, as one-parameter models we will adopt the SIS or constant-\ML\ profiles (e.g. \citealt{Tortora+09,SPIDER-VI}). As a multi-parameter profile we will use NFW+S\'ersic for DM+stars, where we fix the NFW dark mass parameters by using independent information from the literature, e.g. the concentration and virial mass by adopting realistic \cvir--\Mvir\, and also force the the virial mass to obey an observed virial-to-stellar mass relation, \Mvir--\mst, where the stellar mass, \mst, can be a free parameter (the impact of these assumptions have been extensively discussed in \citealt{TRN13_SPIDER_IMF,Tortora+14_DMslope}).
In the latter case, this allows us to have freedom on the stellar mass-to-light ratio and evaluate the IMF normalization. In this paper we will adopt as reference results the ones derived in \cite{TRN13_SPIDER_IMF}, where the DM halo is set using the \cvir--\Mvir\ correlation from N-body simulations \citep{Maccio+08} and \Mvir--\mst\ correlations from abundance matching results in
\citet{Moster+10}. This reduction of the parameter space is obliged, in our case, by having a single observable for each galaxy.
However, the robustness of such approaches have been demonstrated in more complex analysis using spatially resolved kinematics (e.g. \citealt{Cappellari+13_ATLAS3D_XV}) and combining dynamics with strong and weak gravitational lensing (e.g. \citealt{Auger+10}).
\item The resulting best-fit mass profile then provides the
total spherical \Mdyn(r).
\end{enumerate}

The dynamical procedure described above does not take into account the contribution of the black hole (BH) mass, orbital anisotropy and rotation. We have estimated that these latter, in most of the cases, do not impact the \Mdyn(r) estimates for more than few per cents (see more in
\citealt{Tortora+09, SPIDER-VI,Tortora+18_KiDS_DMevol}).

\vspace{0.3cm}

\subsection{Central dark matter}\label{subsec:DM-fraction}
The main prediction of the virial formula (Eq. \ref{eq:M_virial}) and Jeans analysis (\Sec\ref{subsec:Jeans}) is \Mdyn(r),
which, under the reasonable assumption of no cold gas and dust in the galaxy centers, is made up only from the stellar mass, $M_*$ and the dark mass, $M_{\rm DM}$, at a given radius $r$. Hence, if an accurate and independent estimate of $M_*(r)$ is available, one can quantify the dark mass just by $M_{\rm DM}(r)=\Mdyn(r)-M_*(r)$ in the galaxy regions probed by the observed $\sigma$. Thus, the DM content can be characterized by computing
the following quantities. First, the
de-projected DM fraction, $\fdm(r) = 1 - \mst(r) / \Mdyn(r)$ or, equivalently, the total-to-stellar mass ratio $\Mdyn(r)/\mst(r)$, which avoids the negative values
arising in the DM fractions when $\mst(r) > \Mdyn(r)$. Second, the de-projected average DM density, $\rhoDM =
( \Mdyn(r) - \mst(r) ) / (\frac{4}{3}\pi r^{3})$, to probe the local average density
of DM, indipendently of the slope of the mass density profile from the center to $r$. Hereafter,  we
will also refer to $\Mdyn/\mst(r=\Re)$ ($\langle \rho_{DM}(r=\Re)
\rangle$), i.e. the DM fraction (density) computed within a
de-projected radius equal to the projected \Re, as the ``central''
DM fraction (density).\\

\subsection{Datasamples}\label{subsec:data}

In order to apply the method illustrated in \Sec\ref{subsec:Jeans} and derive the quantities in \Sec\ref{subsec:DM-fraction}, one need datasets for which accurate surface photometry  (e.g. high quality imaging in one or more optical and/or near-infrared, NIR, bands) and accurate stellar masses (e.g from optical + NIR multi-band stellar population synthesis models or from spectroscopy) are available. In case one wants to study the evolution with redshift of the DM content, these datasets should provide uniform measurements from $z=0$ to some higher redshits. For our analysis we rely on two datasamples with these properties.

Our reference sample is made of massive and red galaxies collected in
\cite{Tortora+18_KiDS_DMevol}, including stellar masses and
structural parameters derived from Kilo Degree Survey (KiDS)
imaging (\citealt{deJong+15_KiDS_paperI,deJong+17_KiDS_DR3}).
This sample encompasses a quite large range of redshifts and it is suitable to study the dark and luminous properties of galaxies from $z\sim0.1$ to $z\sim 0.7$. The dataset consists of
$\sim 9700$ selected ETGs with optical and NIR photometry information. The sample is complete in stellar mass, obtained by the SED fitting of the KiDS optical bands and assuming a \cite{Chabrier01}, at $\log \mst/\Msun > 11.2$ and redshifts $z < 0.7$. Structural parameters (e.g., effective radius \Re\ and S\'ersic
index $n$) are obtained by a PSF-convolved S\'ersic fit of the KiDS galaxy images in g-, r- and i-bands (using 2DPHOT, \citealt{LaBarbera_08_2DPHOT}). In our calculations, these are rest-framed, as described in \cite{Tortora+18_KiDS_DMevol}.
Spectroscopic redshifts and central velocity dispersions are taken from
SDSS--DR7 ($z<0.2$ \citealt{SDSS_DR7_Abazajian}) and BOSS
($z>0.2$, \citealt{Thomas+13_BOSS}). We refer the reader to \cite{Tortora+18_KiDS_DMevol} for more details about data selection and analysis.

We also use an external properly selected sample of local galaxies with analogous photometry and spectroscopy information, as control sample at $z=0$. This is made of $\sim 4300$ giant ETGs
drawn from the SPIDER project (\citealt{SPIDER-I,SPIDER-VI}). It includes stellar masses derived from SED fitting their optical and
NIR photometry (\citealt{SPIDER-V}) using a
\cite{Chabrier01} IMF. This dataset also includes
galaxy structural parameters, derived
from $g$ through $K$ wavebands, and the SDSS central-aperture
velocity dispersions, SPIDER ETGs are defined as luminous
bulge-dominated systems, featuring passive spectra in the central
SDSS fibre aperture (\citealt{SPIDER-I}).

There are other datasets that have been used to perform similar studies. In the following, we will compare our results with this independent
literature, when available.\\

\section{Correlation with structural parameters and mass
probes}\label{sec:DM_correlations}

\vspace{0.2cm}

To characterize the central DM content in massive ETGs, we start
looking for correlations among \TtoSM\ and galaxy parameters.\\

\subsection{Results for KiDS and SPIDER datasamples}

\vspace{0.2cm}

\begin{figure*}
\begin{center}
\includegraphics[width=17.5cm]{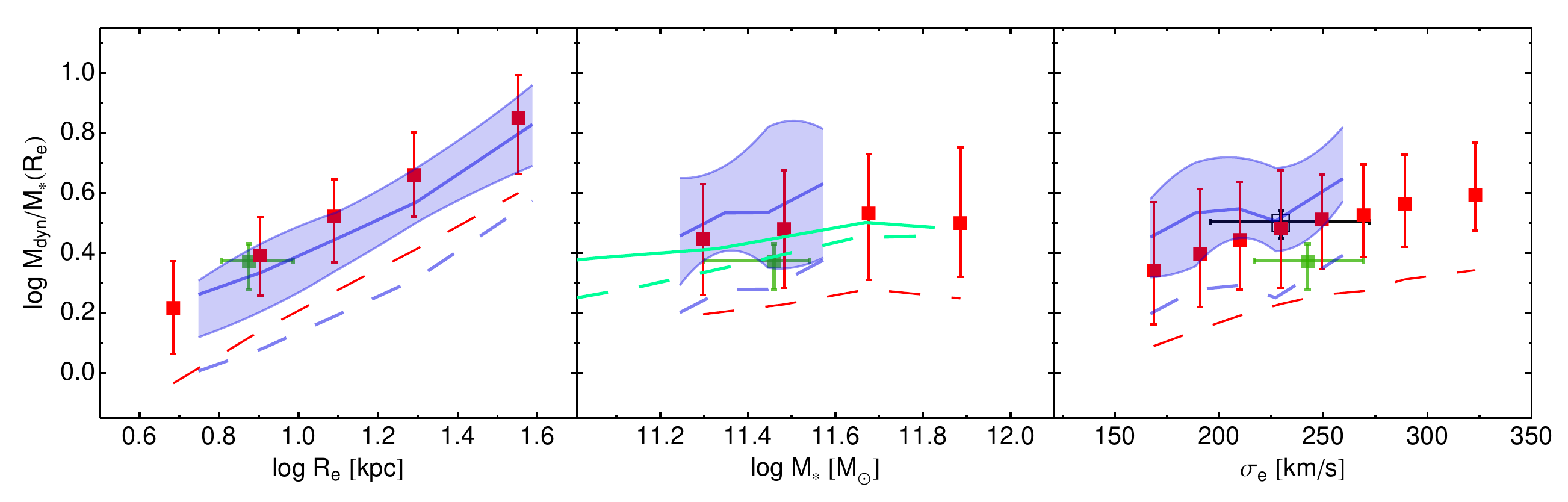}
\caption{
The total-to-stellar mass ratio \TtoSM\ within effective radius, \Re, assuming a Chabrier IMF, as a function of \Re\ (left panel), total stellar mass \mst\ (middle panel) and velocity dispersion within effective radius
$\sigma_{\rm e}$ (right panel). Lines correspond to medians, and shaded regions or error bars to 25--75th percentile. Red squares with error bars are for the whole KiDS sample. Dark blue line and light blue region are for SPIDER galaxies with $\mst > 10^{11.2}\, \rm
\Msun$. Red and blue dashed lines are medians for KiDS and SPIDER galaxies adopting a Salpeter IMF.
Green squares and error bars are for SLACS lenses from \citet{Auger+10_SLACSX}. The cyan lines in the middle panel are from IllustrisTNG simulated galaxies in \citet{Lovell+18_Illustris}: solid line is from the reference IllustrisTNG, while dashed one is for galaxies which, simulated within the full-physics IllustrisTNG simulation, are placed in their corresponding DM haloes simulated within the DM-only simulation. In the right panel, black open square and error bars are for the results obtained
applying a Schwarzschild's orbit superposition technique to ETGs in \citet{ThomasJ+11}.}\label{fig:DM}
\end{center}
\end{figure*}

We will consider first the results found using the KiDS dataset and compare these with results obtained on the SPIDER dataset at $z\sim0$.
In \Fig\ref{fig:DM} we show \TtoSM\ as a function of effective radius, stellar mass and velocity dispersion, We fix the IMF to the Chabrier one, which implies that the measured \TtoSM\
trends correspond to a variation in the DM content. Red symbols represent the estimates of the KiDS sample with redshift $z < 0.7$ and masses $\log
\mst/\Msun > 11.2$. Dashed blue lines with light blue shaded
regions represent the SPIDER sample, assuming $g$-band structural
parameters. Error bars and the shaded regions are the 25--75th per
cent quantiles. All the correlations are significant at more than
99 per cent. The best-fitted slopes of the correlations are also
reported.

\Fig\ref{fig:DM} shows a tight and positive correlation between (the logartithm of) \TtoSM\ and \Re\ with a slope
$\alpha=0.72$. This can be interpreted as a
physical aperture effect, where a larger \Re\ subtends a larger
portion of a galaxy DM halo (e.g. \citealt{NRT10,Napolitano+11_PNS}). In practice, larger \Re\ are found in galaxies with larger stellar mass; however, being these massive ETGs characterized by a steep halo-to-stellar mass relation (e.g. \citealt{Moster+10}), the halo DM mass is increasing too. The net effect is the observed positive correlation between \TtoSM\ and \Re . A similar correlation can also be found for the S\'ersic index (see \citealt{Tortora+18_KiDS_DMevol}), due to the positive correlation existing between the $n-$index and effective radius (e.g. \citealt{SPIDER-VI}). However, if we plot \TtoSM\ as a function of \Re\ and bin the galaxies in terms of the S\'ersic index, then we see that there is a negligible dependence on $n$ (this result is not shown in the figure for brevity). Because smaller effective radii correspond to higher stellar densities, this correlation with \Re\ also translates to a sharp anti-correlation between DM
content and central average stellar density,
which has been reported for the first time in \cite{SPIDER-VI} and
subsequently confirmed in \cite{Tortora+18_KiDS_DMevol}. Galaxies with the
smallest \Re\ ($\sim 5 \, \rm kpc$) have the smallest DM fraction ($\sim 35$ per cent), while the largest galaxies ($\Re \sim 35 \, \rm kpc$) present the largest DM content ($\sim
85$ per cent).

Moving to the correlations of the \TtoSM\ with ``mass'' parameters, we find a shallow correlation with \mst, with an average $\TtoSM \sim 3$ (i.e. $67$ per cent of DM) and slope of $0.11$. We also find that \TtoSM\ correlates with \sige\ ($\TtoSM \propto
\sige^{0.89}$). It can be shown that the strong correlations with \Re\ and those with \sige\ translates to a strong positive correlation with \Mdyn\ (\citealt{Tortora+18_KiDS_DMevol}).

The impact of the IMF is also shown: \TtoSM\ for a Salpeter IMF are plotted as dashed lines for both the KiDS and SPIDER samples, obtained by multiplying the stellar mass by a factor of 1.8 (\citealt{Tortora+09}). The change of IMF move downward the \TtoSM\ of a factor $\sim 0.25$ dex. On average, the DM fractions are positive in these very massive galaxies, with \fdm\ close to 0 only in the smallest galaxies.

These trends are consistent with the ones found at
$z\sim 0$ using SPIDER galaxies (blue lines with shaded regions;
\citealt{SPIDER-VI}), also confirming previous results reported in literature
at $z \sim 0$ (e.g., \citealt{Cappellari+06}; \citealt{HB09_curv};
\citealt{Tortora+09,SPIDER-VI}; \citealt{NRT10};
\citealt{Nigoche-Netro+16}; \citealt{Lovell+18_Illustris}), or at
intermediate redshift (\citealt{Tortora+10lensing};
\citealt{Auger+10_SLACSX}; \citealt{Tortora+14_DMevol}). The evolution with redshifts will be discussed in the next Sections.

\begin{figure*}
\begin{center}
\includegraphics[width=14cm]{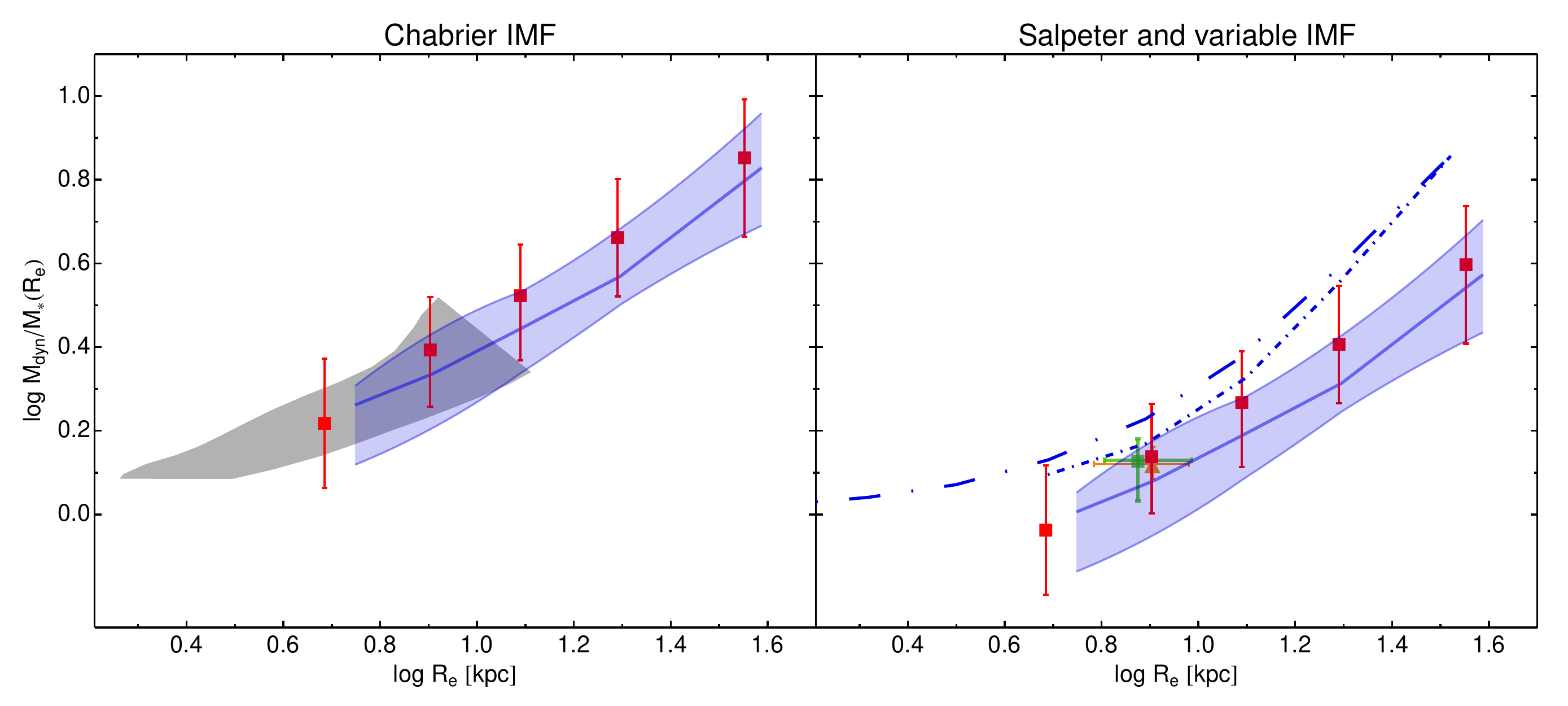}
\caption{The total-to-stellar mass ratio $\Mdyn/\mst$ within
rest-frame effective radius, \Re, as a
function of rest-frame effective radius \Re, assuming a Chabrier IMF (left panel) and Salpeter and variable IMF (right panel). Red and blue symbols are for KiDS and SPIDER, respectively. Following \Fig\ref{fig:DM}, red squares with errors bars and solid lines with shaded regions adopt a fixed IMF (Chabrier and Salpeter in the two panels). Short-dashed-dotted is for IMF variable from
\cite{TRN13_SPIDER_IMF} and \cite{Tortora+14_DMslope} with $\log
\mst/\Msun > 11.2$, while long-dashed-dotted with $\log \mst/\Msun
> 10.5$. The gray region
is from Illustris TNG (\citealt{WangY+20_IllustrisTNG}), extracted
from Figure 15 in \citet{Shajib+21}; note that in that figure they
only show results for $\Re \lsim 10 \, \rm kpc$.} Orange triangle
is from the lensing analysis in \cite{Shajib+21}, in a good
agreement with \cite{Auger+10_SLACSX} results, plotted in
green.\label{fig:DM_vs_Re_lit}
\end{center}
\end{figure*}
~\\

\subsection{Comparison with the literature}

We compare our results with \TtoSM\ estimates from
gravitational lensing of the SLACS sample with
$\log \mst/\Msun > 11.2$ and average redshift of $z \sim 0.2$
(\citealt{Auger+09_SLACSIX,Auger+10_SLACSX}). For a fair comparison, we have homogenized the lensing results, and derived medians and 25--75th percentiles. These are shown in Figure \ref{fig:DM} with green
symbols. A good agreement is clearly seen for the
\TtoSM--\Re\ correlation, while we notice that at fixed
\mst\ and  \sige\, SLACS \TtoSM\ are smaller of $\sim 0.1-0.2$
dex than the median KIDS relation.
However, we need to remark that, at
fixed \mst, the SLACS sizes are smaller than the ones of the KiDS
sample by $\sim 0.15$ dex, while velocity dispersion are $\sim 0.03$ dex higher, which implies than that SLACS
\TtoSM\ are smaller of $\sim 0.1$ dex within their \Re. The
smaller sizes of SLACS galaxies are also clear from the
\TtoSM--\Re\ correlation, where SLACS
galaxies have sizes concentrated towards smaller values, with
respect to the range of sizes of SPIDER and KiDS datasamples (see also \citealt{Tortora+18_KiDS_DMevol}).

In the middle panel of \Fig\ref{fig:DM} we also compare our \TtoSM\ with the results from the state-of-the-art IllustrisTNG simulations from \citet{Lovell+18_Illustris}, which adopt a Chabrier IMF (cyan lines). The agreement with both SPIDER and KiDS galaxies is quite good, considered the uncertainties in observations and simulations.

In the right panel of the same figure we also show that the \TtoSM--\sige\ correlation for KiDS galaxies, and in particular for the SPIDER sample, is fairly consistent with the DM fractions from
Schwarzschild's orbit superposition models in axisymmetric
potentials in \cite{ThomasJ+11} applied to a sample of 16 COMA ETGs.

In Figure \ref{fig:DM_vs_Re_lit} we expand the comparison of the \TtoSM--\Re\  with further theoretical and observational
results, and investigate the impact of a non-Universal IMF. In particular, in the left
panel we confirm one of the results presented in \Fig\ref{fig:DM} thanks to the comparison with \cite{Lovell+18_Illustris}: our DM fractions are fairly well consistent
with the expectations from \cite{WangY+20_IllustrisTNG}, who also use IllustrisTNG.
IllustrisTNG sample also includes galaxies with $\log\mst/\Msun <
11.2$ and have smaller sizes, for this reason we find an overlap
with the smallest galaxies in our sample. This result provide a further confirmation
that the prescriptions adopted by IllustrisTNG are able to
realistically provide a quite good agreement with observations.
Moreover, the comparison with state-of-the-art observations is presented in the right-hand panel of \Fig\ref{fig:DM_vs_Re_lit}, where
we also compare with the recent results from \cite{Shajib+21}, who
determine the mass density slope and DM fraction of a sample of
SLACS lenses, using their strong lensing data, velocity
dispersions and weak lensing constraints. These results are
plotted as orange symbols. \cite{Shajib+21} show that most of the
totality of SLACS galaxies are well fitted by a NFW profile with a
Salpeter IMF. For this reason, we converted our \TtoSM\ to
Salpeter-based versions in this panel. As a consistency check we also overplot
the results from \cite{Auger+10_SLACSX}, using the same IMF and find a striking consistency of the two analyses.\\

\vspace{0.2cm}

\subsection{The intruder: the impact of Initial Mass Function}

\vspace{0.2cm}

In \Fig\ref{fig:DM} we have shown that, on average the DM fractions are positive, independently of the IMF adopted. However the $\fdm$ estimates present single cases where galaxies have negative $\fdm < 0$, implying unphysical $\Mdyn(\Re) < \mst(\Re)$. In the KiDS dataset, only the $\sim 6$ per cent of the galaxies show negative DM fractions, if a Chabrier IMF is adopted. On the other hand, using a Salpeter IMF, produce, by definition,
smaller DM fraction, and for $\sim 23$ per cent even negative (unphysical) values.
The fraction of such negative \fdm\  is larger at higher redshifts: this means that if the Salpeter IMF is an inappropriate choice, this is even worse at higher redshift. This is a well known
critical effect that has been discussed in previous works
(see e.g. \citealt{Tortora+09}; \citealt{NRT10}; \citealt{SPIDER-VI}). Despite a significant fraction of these
negative \fdm\ is, in principle, compatible with observational scatter in
\mst\ and \Mdyn\ (see \citealt{NRT10}), in order to avoid unphysical results, one is left with no
complete freedom on the assumption of the IMF to adopt. In particular, a higher stellar \ML\ normalization is unphysical for
those systems which tend to have smaller \fdm\ (e.g. the ones with
smaller sizes and dynamical masses, larger stellar densities,
higher redshift, etc.).

These indications cope with a large number of observational studies that in the last decade have suggested that IMF in ETGs is different from the one estimated by star counts in our Milky Way, where the standard forms for the IMF were identified (e.g.
\citealt{Kroupa01}; \citealt{Chabrier03}).
Now, this hypothesis of IMF universality is questioned by different lines of observational evidence, using completely
independent data, as spectral features, galaxy dynamics and
gravitational lensing (e.g., \citealt{Treu+10};
\citealt{Conroy_vanDokkum12b}; \citealt{Cappellari+12,
Cappellari+13_ATLAS3D_XX}; \citealt{Spiniello+12};
\citealt{Goudfrooij_Kruijssen13};
\citealt{LaBarbera+13_SPIDERVIII_IMF};
\citealt{TRN13_SPIDER_IMF,Tortora+14_DMslope,Tortora+14_MOND,Tortora+18_Verlinde};
\citealt{McDermid+14_IMF};
\citealt{Martin-Navarro+15_IMF_variation}; \citealt{Corsini+17};
\citealt{Li+17_IMF}). Among the others, using SPIDER data, we have
found a larger stellar mass in the most massive galaxies (high
velocity dispersion) than that provided by a Milky-Way Chabrier
IMF, that can be translated to a bottom-heavy IMF (Salpeter-like)
or a larger dwarf-to-giant star ratio (\citealt{TRN13_SPIDER_IMF}). On
the contrary, in ETGs with a low-velocity dispersion, the IMF
resembles the one that is found in the Milky-Way. We have found
that this very tight correlation with velocity dispersion is safe
independently of the galaxy model adopted, and also in alternative
gravity scenario, in which DM is not included, as MOND
(\citealt{Tortora+14_MOND}) and Emergent Gravity
(\citealt{Tortora+18_Verlinde}). Despite these positive strong
trends with $\sigma$ and stellar mass (when this last takes into account of the dynamically inferred IMF), milder trends are found with \Re, n and \mst\ when
a fixed IMF is adopted for this latter
(\citealt{Tortora+14_DMslope}).

If, a strongly varying IMF as a function of velocity dispersion is
translated into a central DM fraction, we find a fairly universal
DM fraction, consistent with $\fdm\sim0.2$ for a NFW profile DM halo\footnote{We remind the reader that we fix the DM NFW halo using the \cvir--\Mvir\ correlation from N-body simulations in \cite{Maccio+08} and \Mvir--\mst\ correlations from abundance matching results in
\citet{Moster+10}, and finally use the Jeans equation to constrain the IMF normalization.} with no further baryonic effects (e.g. adiabatic contraction, AC, \citealt{Gnedin+04}), or $\sim0.4$
if one accounts the physics of baryon collapse, e.g. via AC (\citealt{TRN13_SPIDER_IMF}). The
positive correlation of DM fraction with mass has been typically
invoked as the driver for the ``tilt'' of the ETG fundamental
plane (\citealt{Tortora+09}), but it turns out that this latter can also be explained, also if partially, by a realistic DM halo model and a non-universal IMF.

{\it Here, for the first time, in the
right-panel of \Fig\ref{fig:DM_vs_Re_lit}, we can show that, instead,
the positive trend between \TtoSM\ and \Re\  still survives if a
non-universal IMF is adopted.} And this is expected considering
that, unlike the trend with $\sigma$, IMF varies mildly with
\Re. Short-dashed-dotted is for IMF variable from
\cite{TRN13_SPIDER_IMF} and \cite{Tortora+14_DMslope} with $\log
\mst/\Msun > 11.2$, while long-dashed-dotted with $\log \mst/\Msun
> 10.5$. The \TtoSM\ values found in this way are slightly larger
than those made with SPIDER and KiDS, and the results from
lensing; however, at least a part of these discrepancies can be also ascribed
by the fact that K-band is used, and thus using g-band the curves should move towards right.

\begin{figure*}
\begin{center}
\includegraphics[width=14cm]{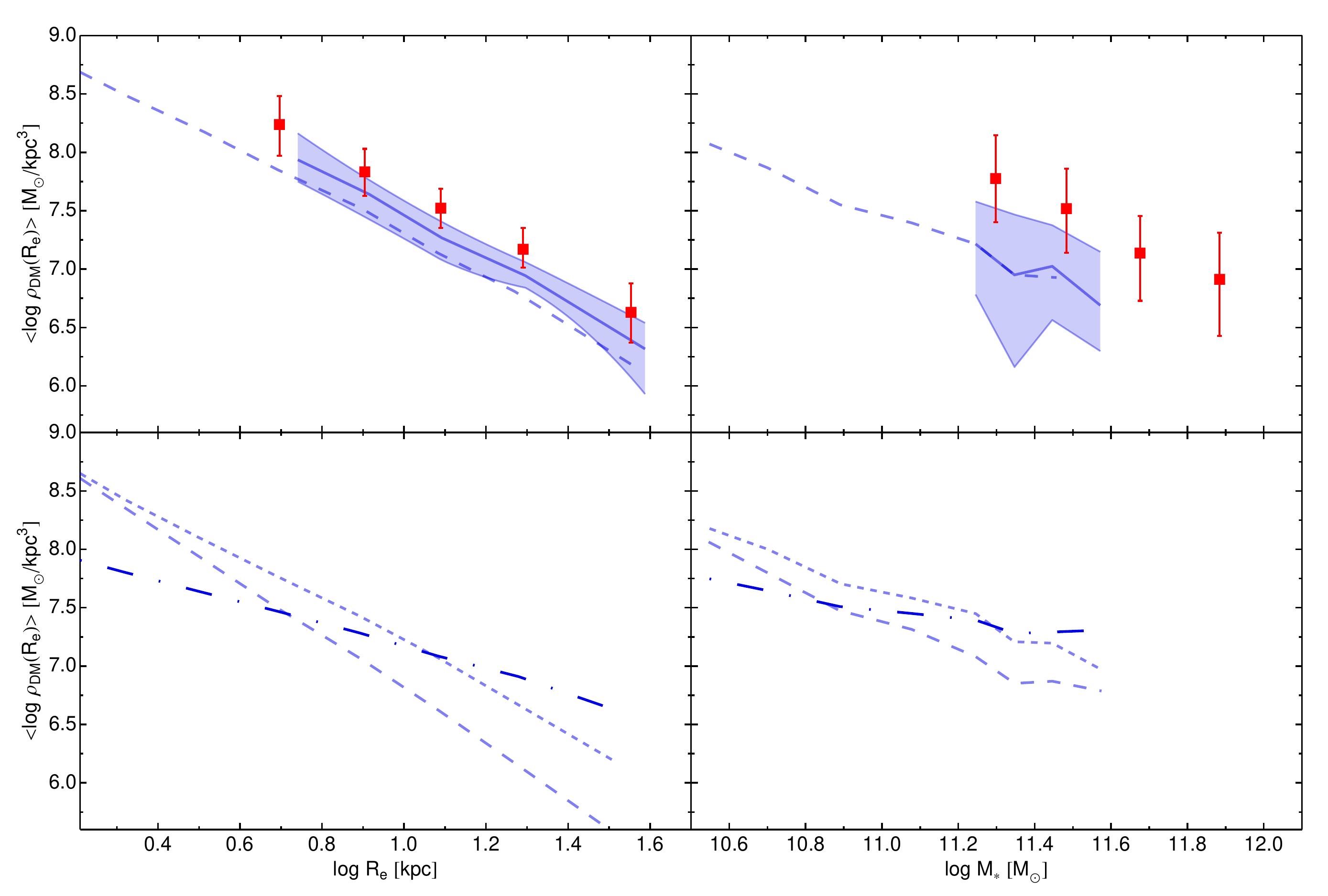}
\caption{Average (log) DM density as a function of \Re\ (left panels)
and \mst\ (right panel). In the top panels KiDS data in red are
compared with SPIDER data in the same mass range. The dashed lines
are plotted considering SPIDER galaxies with $\log \mst/\Msun
> 10.5$ dex. g-band rest-frame structural parameters are used. In the bottom panel, instead, we focus on the SPIDER
datasets. K-band structural parameters are used. Short- and
long-dashed lines are the medians for a SIS and a constant-\ML.
The point-dashed line is the result obtained fitting a NFW for DM
halo and a free IMF.}\label{fig:rhoDM}
\end{center}
\end{figure*}

\section{DM density, DM and total mass density slope}\label{sec:av_DM}
Once deduced the DM content on the central regions of galaxies, one can check how this connects to the whole DM halo properties, e.g. as predicted from the cosmological simulations. This step is tricky because obtaining inferences on the total DM halo from the measurements in the center is impracticable. The only possible way is to assume some standard recipe for the DM halo and try to match the predictions from these recipes with the central DM inferences (see e.g.  \citealt{Tortora+09,SPIDER-VI,Tortora+14_DMevol}; \citealt{NRT10}).

There are different
mechanisms that can affect the central DM content and decouple it
from the overall DM. In fact, the central \fdm\ can also reflect the
local conditions and thus the environment density at the time of
initial halo collapse. Moreover, it is well known that baryons
interact with DM, changing its distribution (as in the case of
AC, \citealt{Gnedin+04}). From a practical
point of view, the DM fraction is somewhat more strongly dependent on the particular
values of \Re\ for the stars rather than the DM
properties directly. These effects imply strong degeneracies among the parameters in the galaxy models (see e.g. \citealt{NRT10}), and providing a quantitative comparison to cosmological theory is a necessary but difficult task.

One direct way to investigate the DM halos and study their properties is to define its average density within some small radius, i.e., \rhoDM.
Following \cite{NRT10} and \cite{Tortora+10lensing}, we focus on
correlations of the average DM density with galaxy size and
stellar mass for SPIDER and KiDS datasamples. \Fig\ref{fig:rhoDM}
shows that $\langle\rho_{\rm DM}\rangle$ strongly anti-correlates
with \Re. Considering the aperture effect and assuming DM halo
homogeneity, this implies that we are measuring a mean DM density
profile with radius. In fact, as firstly discussed in \cite{NRT10}, if we assume a power-law for the DM profile, i.e. $\rho_{\rm DM} (r) \propto r^{\alpha}$, with $\alpha$ negative, then, with a little of algebra, for $\alpha > - 3$ one finds $M_{\rm DM} \propto r^{\alpha + 3}$ and finally $\rhoDM(r=\Re) \propto \Re^{\alpha}$. Hence, the trend of $\rhoDM(\Re)$ with \Re, calculated on a sample of galaxies, provides the average slope of the DM profile in the central regions of that galaxy sample.
For an NFW halo, near \Re\ is $\alpha \sim -1.1$, $-1.3$. AC makes the DM cuspier, with $\alpha \sim - 1.6$, $-1.9$. For KiDS and SPIDER, the slope of this correlation is $\sim -1.8,
-1.9$, consistently with what originally shown by \citet{NRT10}, using a completely different datasample. Therefore, this steep slope can be indicative of cuspier-than-NFW halos, perhaps as induced by AC. But this result can be driven by the choice of a Chabrier IMF. Hence, we cannot exclude that a standard cuspy
NFW model and a larger amount of stars due to a Salpeter IMF (e.g., \citealt{Shajib+21}). We have verified that using an alternative
constant-M/L profile yields similar results still fully consistent
with a cuspy halo (see also a detailed discussion in \citealt{NRT10}).

A similar negative correlation is also obtained when the average
densities are plotted as a function of stellar mass. The most
massive galaxies are characterized by systematically lower values
of the DM density; in fact, at increasingly higher masses, size
get bigger and bigger, lowering the DM average density. In both
the panels, the more distant galaxies in the KiDS sample have, at
fixed \Re\ or \mst\ larger average DM densities, also because of
the systematically smaller sizes in these galaxies. The results
for SPIDER galaxies are almost unchanged if the lower-mass cut of
$\log \mst/\Msun = 11.2$ dex is removed (dashed lines).

In the bottom panels of \Fig\ref{fig:rhoDM} we focus on the local SPIDER
datasample and investigate the impact of different assumptions for
galaxy model. Differently from the previous plots, we use here the
K-band structural parameters to calculate the DM fractions.
Consistently with the results in \cite{Tortora+09} and
\cite{SPIDER-VI}, we find that if we model the total mass
distribution with a constant-\ML\ (long-dashed line), then the DM
content is decreased with respect to an isothermal profile
(short-dashed line). We notice that the trend with radius is steeper than the SIS profile  when the constant-\ML\ is adopted, pointing
to a slope of $\sim -2.3$.

We can now do a step forward, by comparing these correlations with the average DM densities obtained fitting a NFW. We recall that, in this case, we leave the IMF free to change, and assume a realistic \cvir--\Mvir\ and \Mvir--\mst\ correlations, to constrain the shape of the total mass profile at $\lsim \Re$.
In this case we find that in order to match the DM distribution of a NFW
profile the total mass profiles of the most massive galaxies are well reproduced by an
isothermal profile, while lower mass galaxies are better fitted by
a constant-\ML, this implying a steeper mass density in the central regions.
This result implies a {\it DM non-homology in the central galaxy regions, i.e. the total mass density slope in ETGs is not universal and is a function of the stellar mass}, a result that we had suggested in \cite{Tortora+09}, and demonstrated in more details in \cite{Tortora+14_DMslope}. These results have been confirmed by independent observations and simulations (e.g., \citealt{Remus+13}, \citealt{Dutton_Treu14}, \citealt{WangY+20_IllustrisTNG}). The steepest slopes are found at $\mst\sim3 \times 10^{10}\, \rm
\Msun$, while isothermal profiles are common in the most massive
galaxies. This trend with mass may be related to a varying role of
dissipation and galaxy mergers with galaxy mass (see also
\citealt{Tortora+19_LTGs_DM_and_slopes}).\\

\section{Evolution with redshift}\label{sec:DM_evolution}

One firm evidence of the hierarchical scenario in a $\Lambda CDM$ cosmology is the size and mass growth of galaxies with time, after the Big Bang
(\citealt{Daddi+05}; \citealt{Trujillo+06,Trujillo+07,Trujillo+11};
\citealt{Saglia+10}; \citealt{Tortora+14_DMevol,Tortora+18_KiDS_DMevol}, \citealt{Roy+18}). These observations all seem to exclude
a simple monolithic-like scenario, according to which the bulk of the stars is
formed in a single dissipative event followed by a passive
evolution, while they seem to support a scenario where galaxy evolution is mainly driven by merging, even though of different kind.

While the understanding of the stellar component of galaxies has been supported by a multiplicity of observational probes, basically measuring the properties of the light distribution in galaxies (e.g., \citealt{SPIDER-I}, \citealt{Tortora+16_compacts_KiDS,Tortora+18_UCMGs}; \citealt{Roy+18} and reference therein) and their stellar populations (e.g., \citealt{Kauffmann+03}; \citealt{Gallazzi+05}; \citealt{Renzini06}; \citealt{Sanchez+12_CALIFA_I}; \citealt{Maraston+13_BOSS}), evidences about the growth of the DM with time come mainly from theory (e.g., \citealt{CW09}; \citealt{Moster+10}; \citealt{Lovell+18_Illustris}) but there are yet little analyses trying to systematically study the evolution of size and mass of DM halo
with redshift.\\

\vspace{0.2cm}

\begin{figure*}
\centering
\includegraphics[trim= 0mm 0mm 0mm 0mm, width=0.99\textwidth,clip]{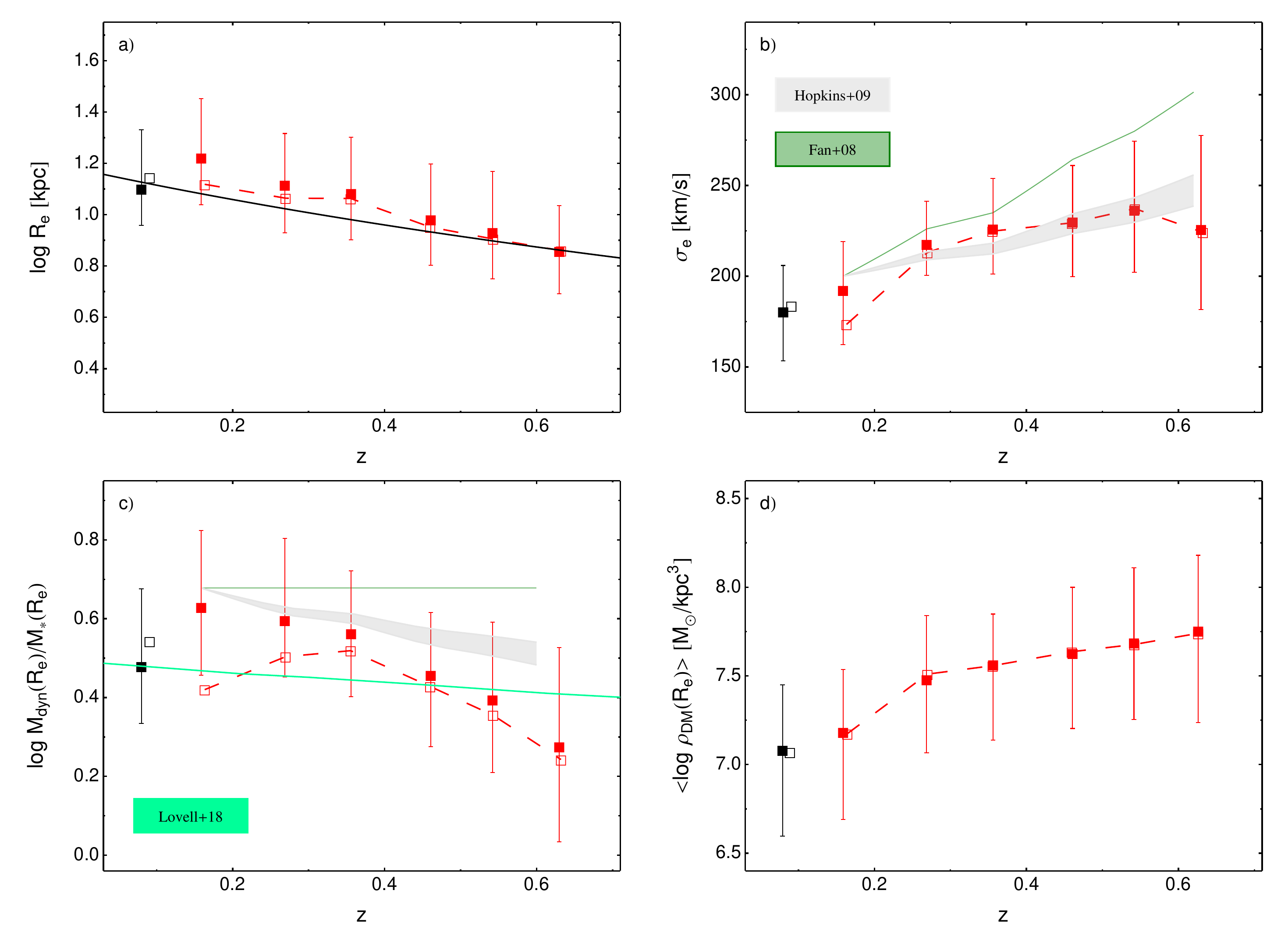}
\caption{Evolution with redshift of (log) \Re\ (panel a), \sige\
(panel b), (log) \TtoSM\ (panel c) and (log) \rhoDM\ (panel d) for $\log \mst/\Msun >
11.2$. Red
symbols are as in previous figures. Open black square with error
bar is median and 25--75th percentiles for SPIDER galaxies, while
open black square is the median for the SPIDER sample when the
progenitor bias is taken into account. The black solid line in the top panel is taken
from the average \Re/$R_{\rm e,0}$--z trends (with $R_{\rm e,0} =
\Re(z\sim 0)$) for spheroid-like galaxies in
\citet{Trujillo+07} normalized to $R_{\rm e,0} = 15\, \rm kpc$. Shaded gray region and green line are our
predictions using the merging model of
\citet{Hopkins+09_DELGN_IV} and the "puffing-up" scenario
from \citet{Fan+08}, respectively. The cyan line in the bottom-left panel is for galaxies with a stellar mass of $\sim 10^{11} \, \rm M_{\odot}$ from IllustrisTNG simulations (\citealt{Lovell+18_Illustris}).
See explanations in the main text.}\label{fig:evolution}
\end{figure*}

\subsection{Evolution of the dark matter fractions at fixed mass}\label{subsec:evol_fix_mass}

In the previous paragraphs we have seen that the \TtoSM, as well as the \fdm, are tightly correlated to observables like the velocity dispersion and the (stellar or dark) masses. When studying the evolution of the galaxy DM fractions with redshift, one needs to estimate how much of this evolution is driven by the correlated quantities and how much by the actual mass assembly of galaxies. In \Fig\ref{fig:evolution} we show the dependence on redshift of \Re, \sige, \TtoSM\ and \rhoDM, at fixed stellar mass, using the KiDS sample as
reference\footnote{We refer
the reader to \cite{Tortora+18_KiDS_DMevol} for the
analysis of systematics and for a full description of data selection and analysis.}.
With respect to \cite{Tortora+18_KiDS_DMevol}, we show here the results for all the galaxies with $\log \mst/\Msun > 11.2$. First, we clearly see the effect of the size growth of galaxies, as the galaxy effective radii are smaller at earlier epochs and increase toward $z=0$ (\citealt{Daddi+05}; \citealt{Trujillo+06,Trujillo+07};
\citealt{Buitrago+08}; \citealt{vanderWel+08}\; see also \citealt{Roy+18}, for further
details about size evolution in KiDS galaxies). To quantify this size-redshift relation, we can use a standard formula $\Re =
R_{\rm e,0}(1+z)^{\alpha}$ to fit the data. In
the case of no progenitor bias correction (i.e., assuming that the same kind of galaxies have evolved from one redshift bin to another, red filled squares with
error bars), we find a slope, $\alpha = -2.4$. If the progenitor bias is taken into account (i.e. galaxies observed today as passive systems might be active in earlier redshift bin, open
squares with dashed red lines), the slope becomes shallower, i.e.
$\alpha = -1.8$. This is because active/disk systems tend to have larger sizes than passive/spheroids systems at high$-z$, hence increasing the overall sizes at high$-z$ and diluting the size-redshift relation.
Overall, these measured slopes are steeper than other literature results for spheroid-like systems with $\mst > 10^{11}\,
\rm \Msun$ (solid black line
in the top panels in \Fig\ref{fig:evolution};
\citealt{Trujillo+07};
\citealt{Conselice14_review}). However, at low$-z$ we find a good agreement
with the \Re s from the local sample from SPIDER.

In the panel b) of \Fig\ref{fig:evolution}, we plot the effective velocity dispersion, \sige, as a
function of the redshift. Here the evolution with redshift
looks shallower than the one shown by the \Re, with higher$-z$ galaxies having slightly larger
velocity dispersions than the lower$-z$ ones (\citealt{Cenarro_Trujillo09};
\citealt{Posti+14}). We can quantify the \sige$-z$ evolution using the relation $\sige = \sige_{\rm
,0}(1+z)^{\alpha}$. The estimated slopes are $\alpha = 0.48$ (without progenitor
bias) and $\alpha = 0.73$ (with progenitor bias). In general, we find a good agreement with local
(\citealt{SPIDER-I}; \citealt{SPIDER-VI}), intermediate-z
(\citealt{Beifiori+14}) and higher-z (\citealt{Saglia+10};
\citealt{Tortora+14_DMevol}) measures.

The total-to-stellar mass ratio (assuming a Chabrier IMF) is plotted
in the panel c) of \Fig\ref{fig:evolution}. In this case we see that galaxies show a dominating
DM content in their \Re s at lower redshift (i.e. 75--80\% of DM at $z
\sim 0.2$), while the \fdm\ are smaller at higher--z (40--50\% at $z \sim 0.6$). In this case we can use the relation $\TtoSM = (\TtoSM)_{\rm
0}(1+z)^{\alpha}$ to quantify the dependence of this quantity on the redshift and find $\alpha = -2.4$ (without progenitor bias) and
$\alpha = -1.3$ (with progenitor bias).

Going to the theoretical interpretation of these observed trends, we can compare them with the predictions from two different scenarios, typically adopted to explain the galaxy size
evolution.
First, the merging scenario (MS, hereafter), which predicts that size growth is driven by the accretion of matter as sizes of the merger remnants are larger than those of their progenitors. The merging model of \cite{Hopkins+09_DELGN_IV} also predicts that the velocity dispersion decreases as a consequence of the size growth as the relation $\sigs(z) \propto (1+\gamma)^{-1/2} \sqrt{\gamma +
\Re(0)/\Re(z)}$ holds, where the parameter $\gamma$ sets the DM contribution to the potential relative to that of the baryonic mass.
This $\gamma$ parameter is expected to vary between 1 and 2 (which are the best fitted values for $\mst \sim 10^{11}$ and $\sim 10^{12}\, \rm
\Msun$, respectively).
Second, the "puffing-up" scenario (PS,
hereafter) from \cite{Fan+08}, which predicts that galaxies grow by the effect of quasar feedback, which removes cold gas from the central regions, quenching the star formation and increasing the size of the galaxy. This model predicts also that velocity dispersion increases with increasing redshift as $\sigs(z) \propto \Re^{-1/2}$.

In order to check these predictions against the observed trends we need to derive the expected redshift dependences of velocity dispersions and \TtoSM\ on the redshift, starting from the
\Re--z relation from KiDS median
values in  panels a). We have used the best-fit \Re$-z$ relations discussed above and inserted into the $\sigs(z)$ equation to derive the expected velocity dispersion
in the two schemes. To obtain the \TtoSM(z) relation we need to translate the predicted
velocity dispersions into a \TtoSM. This is done by solving
the spherical Jeans equation (see Eq. \ref{eq:jeans}), which includes 1) the 3D luminous density profile, and 2) the total potential, as a function of redshift.
For the light distribution we have assumed a S\'ersic
profile with $n=4$ for simplicity (i.e. a pure de Vaucouleurs) and the effective radius given by our interpolated $\Re(z)$ relation as defined above. For the total potential we have used the SIS
model. By imposing that the velocity dispersion inferred by the Jeans equation (averaged within \Re) equals the $\sigma(z)$ in the
two scenarios, MS and PS, we can obtain \Mdyn\ and
\TtoSM\ as a function of redshift.
We remark here that in this calculation the relevant information we are interested on is the trend with redshift and not the normalization, which can be adjusted by hand, since in the
$\sigma_{\star}(z)$ formulae the normalization factor is unspecified.

The predicted trends for \sige\ and \TtoSM\ are plotted in the panels b) and c) of \Fig\ref{fig:evolution}.
Here, the PS predicts a too strong evolution (with a variation of $\sim 100\, \rm km \, s^{-1}$
in the redshift window analyzed), which unfits the KiDS results for both the \sige\ and \TtoSM. On the other hand, the
milder evolution predicted from MS matches closer the observations
(\citealt{Cenarro_Trujillo09}). However, even though the agreement with the
velocity dispersion seem very good, the PS model predict a too shallow \TtoSM--z trend with respet to the observed one.

To complement this analysis, we also compare the \TtoSM--z trend with IllustrisTNG expectations from \cite{Lovell+18_Illustris}. Although the agreement with IllustrisTNG is quite good in terms of normalization, as confirmed by the analysis of the trends with \mst\ and \Re\ (\Figs \ref{fig:DM} and \ref{fig:DM_vs_Re_lit}), the simulations predict a change with redshift which is milder than what we find with KiDS data.

For the first time, in this paper we also present the evolution with z of the average DM density. This is shown in panel d) of \Fig\ref{fig:evolution}, where for simplicity we limit to show the results for KiDS and SPIDER datasamples, without comparing with any toy-model. The average DM density within \Re\ reaches its largest values at $z \sim 0.6$ ($<\log \rho_{DM}(\Re)/(\Msun/kpc^{3})>\sim 7.75$ ) and decreases systematically dowm to $<\log \rho_{DM}(\Re)/(\Msun/kpc^{3})>\sim 7$ at the lowest redshifts. This trend is explained by the fact that sizes are larger at smaller redshifts, and the impact on the denominator is stronger than that of the numerator where DM mass is larger at smaller z.\\

\subsection{The evolution of Size-- and Dark matter--mass relations: constraints on merger scenario}\label{subsec:evol_trends}

In the previous section we have found evidence that simple recipes for the size growth and the velocity dispersion evolution in a merging scenario reproduce better the observed evolution of the dark-to-stellar mass ratio, at least for massive ETGs. In this section we want to check further the merging scenario and see whether we can gain more insight on the mechanisms driving the size and mass growth in massive ETGs. In particular, we want to focus on two scaling relations discussed in previous sections, the \Re--\mst\ and
\TtoSM--\mst, and figure how galaxies evolve in this parameter space in response of the joint evolution of size and mass predicted by different kinds of merging events.

Indeed, dissipationless major mergers from
simulations of elliptical galaxies have predicted that the DM
fraction within a certain physical radius decreases mildly after
the merger (\citealt{Boylan-Kolchin+05}). But they have also shown
that the DM fraction within the final \Re\ is greater than the DM
fraction within the initial \Re, because the total mass within
\Re, $M_{\rm tot}(\Re)$, changes, after the merger, more than
$\mst(\Re)$. The problem has been investigated in detail with
hydrodynamic simulations by \cite{Hilz+13} who find that the equal-mass
mergers produce a smaller size increase of multiple minor mergers.
In particular, the variation of \Re\ with respect to the initial
radius, $\Re / R_{\rm 0}$, in terms of the variation of \mst\ with
respect to the initial stellar mass, $\mst / M_{\rm 0}$, is found
to be $\Re/R_{\rm 0} \propto$ $(\mst/M_{\rm 0})^{0.91}$ for the
equal-mass merger and $\propto (\mst/M_{\rm 0})^{2.4}$ for the
minor mergers.

\begin{figure*}
\begin{center}
\includegraphics[width=14cm]{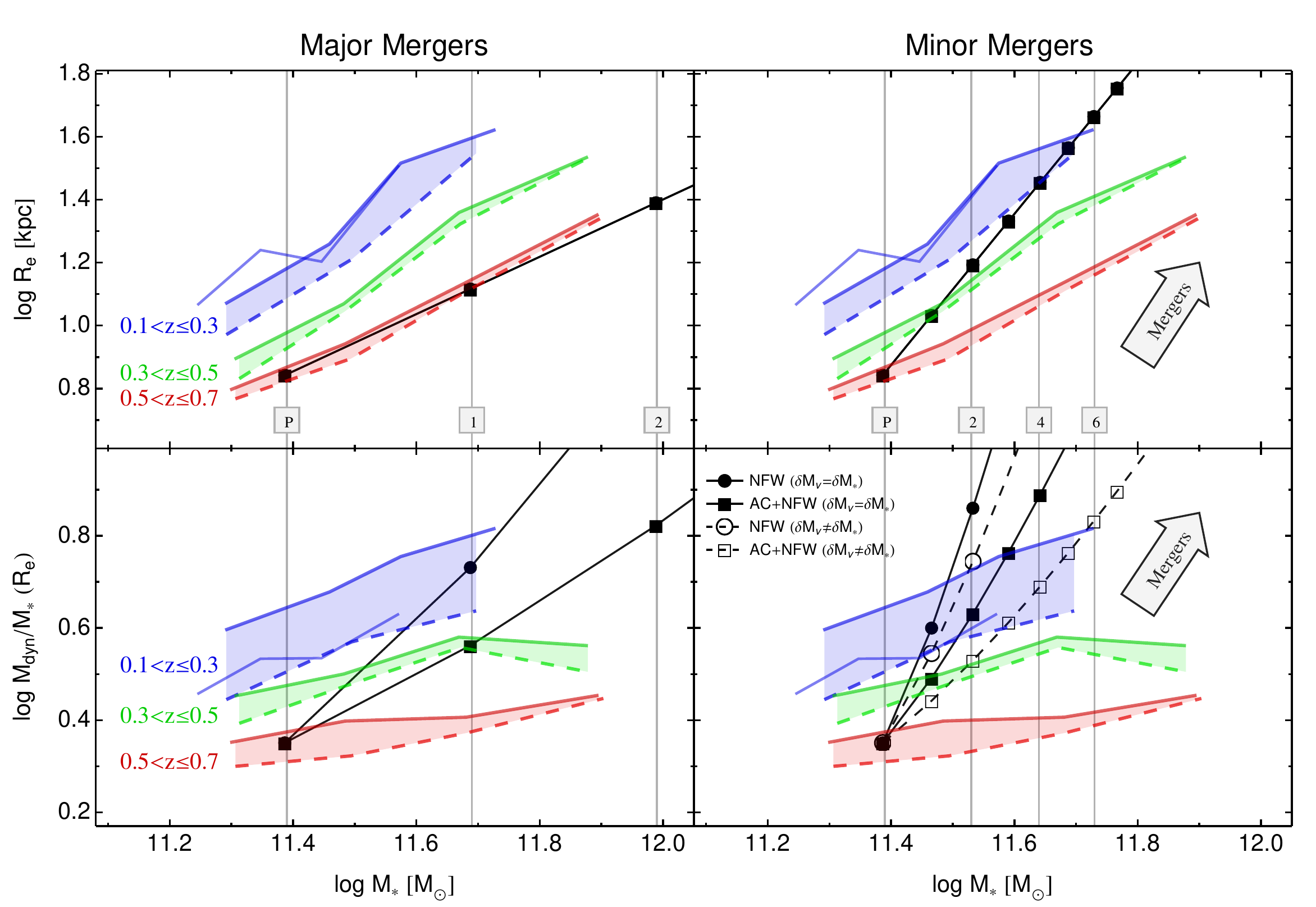}
\caption{Evolution of \Re--\mst\ (top panels) and \TtoSM--\mst\
(bottom panels). Blue, green and red lines are for galaxies in
three redshift bins $0.1 < z \leq 0.3$, $0.3 < z \leq 0.5$ and
$0.5 < z \leq 0.7$. Solid lines are medians for the full sample,
while dashed lines are for results corrected for progenitor bias. The medians for SPIDER galaxies are plotted as thin blue lines.
In the left (right) panels major (minor) toy-model merging predictions (discussed in the text) are drawn in black. We take, as example progenitor galaxy (P), the average galaxy at $\log \mst/\Msun =
11.4$ and evolve it accordingly to the toy-models, the number of mergers is also indicated. In the bottom panels, dots and filled squares
are for NFW and AC+NFW profiles when $\delta \mst = \delta \Mvir$,
and set the \Re, \mst\ and \TtoSM\ after each single merging
event. In the bottom-right panel, open circles and squares are for NFW and AC+NFW
profiles when $\delta \mst \neq \delta \Mvir$. The direction of the merger effect is indicated with arrows in the right panels. Updated and improved version of Fig.~4
of \citet{Tortora+18_KiDS_DMevol}.}\label{fig:evol_toy_models}
\end{center}
\end{figure*}

We can test all these predictions against the \Re--\mst\ and
\TtoSM--\mst\ relations (assuming a fixed Chabrier IMF) for KiDS
galaxies at different redshift, as shown in \Fig\ref{fig:evol_toy_models}. Blue, green and red lines
are for KiDS galaxies in three redshift bins $0.1 < z \leq 0.3$, $0.3 <
z \leq 0.5$ and $0.5 < z \leq 0.7$.
Here we see that lower redshifts galaxies are larger in size and contain more DM in their cores at all values of \mst, confirming the trends in \Fig\ref{fig:evolution}. The trend is weaker if we consider the progenitor bias, which affects mostly the lowest redshift bin (dashed lines in
\Fig\ref{fig:evolution}). We will check if mass, size
and total-to-stellar mass evolution in KiDS galaxies can be
explained, consistently, through major or minor mergers.

To reproduce these trends we need to account for the dark mass inside the effective radius, as we have to reproduce the \TtoSM, where $M_{\rm dyn}=\mst+M_{\rm DM}$ and the evolution of the  $\Re / R_{\rm 0}$ vs. $\mst / M_{\rm 0}$ is incorporated in the relations discussed above.
To do that, we have constructed some simplified toy-models assuming that
$M_{\rm DM} \propto \Mvir r^{\eta}$ around \Re, with $\eta \sim 2$
for a standard NFW and $\eta \sim 1.2$ for a contracted NFW,
hereafter AC+NFW (according with \citealt{Boylan-Kolchin+05}). We start assuming that the systems participating to merging (i.e. the progenitors) all have the same $\Mvir / \mst$, which is reasonable
for most of the stellar mass range covered by our sample, i.e.
$\delta \Mvir \approx \delta \mst$. However, we cannot exclude for the minor merging
case that the virial mass changes at a different rate of the
stellar mass, since the main galaxy is merging with another galaxy
with a different $\Mvir/\mst$.

Having all ingredients set, we have considered the evolution tracks related to the two
different merging types: for the major merging we take the average evolution of \Re\ in terms of \mst\ evolution for the equal-mass merging, i.e. $\Re/R_{\rm 0} \propto (\mst/M_{\rm 0})^{0.91}$; for minor merging we take $\Re/R_{\rm 0} \propto (\mst/M_{\rm 0})^{2.4}$.

In \Fig\ref{fig:evol_toy_models} we show the evolution tracks for a progenitor galaxy with $\log\mst/\Msun = 11.4$. Dots and squares are for NFW and AC+NFW profiles, respectively.
In the left panels, the
major merging tracks are shown as black lines with dots/squares
indicating the events corresponding to masses $\delta \mst \times
M_{\rm \star, 0}$, with the mass increments $\delta \mst =
1,2,4,...$, assuming that $\delta \Mvir = \delta \mst$. The first
dot/square at $\log \mst/\Msun = 11.4$ corresponds to the initial
progenitor galaxy. The second dot/square is the result of one
major merger, which doubles the initial mass of the progenitor,
while the third dot/square corresponds to a second major merger
with mass 4 times the mass of the initial progenitor and 2 times
the mass of the remnant of the first merging event. The minor
merging tracks are shown in the right panels by black
lines. Dots/squares indicate remnants with masses $M_{\rm \star, 0}
+ \delta \mst \times M_{\rm \star, 0}$ where $\delta \mst
=0,0.2,0.4,0.6,...$, and we use two different increment laws for
\Mvir. In the top-right panel we use $\delta \Mvir = \delta \mst$. In the bottom-right panel filled symbols are for $\delta \Mvir = \delta \mst$, while
open symbols correspond to $\delta \Mvir = 0.05 \, \delta \mst$. In
this case we suppose that the initial progenitor galaxy with mass
$M_{\rm \star, 0}$ is experiencing a collection of minor mergers
with galaxies having masses of $20$ per cent of $M_{\rm \star,
0}$.

From these different tracks we can see that major mergers can be excluded, since the predicted evolution
in size of a galaxy in the highest redshift bin (with $z \sim
0.6$) and with stellar mass $10^{11.4}\, \Msun$ is parallel to the
size-mass relation in this same redshift bin (see left panels in
\Fig\ref{fig:evol_toy_models}).

On the contrary, the same galaxy
can evolve to $z \sim 0.2$ experiencing few (5 or 6) minor
mergers, which accrete $\sim 100$ per cent of the initial stellar
mass $M_{\rm \star, 0}$. The example galaxy is not evolving on the top of the $z \sim 0.2$
\Re--\mst\ and \TtoSM--\mst\ relations if we consider that DM and
star accrete at the same rate, for both NFW and AC+NFW (dots and
filled squares in \Fig\ref{fig:evol_toy_models}). The
\TtoSM--\mst\ evolution is too steep, thus after $\sim 2$ minor
mergers the example galaxy would end up on the $z \sim 0.2$
observed \TtoSM--\mst, this number is not consistent with what is
found analyzing the \Re--\mst\ evolution, which would require
$4-6$ mergers to transform the $z \sim 0.6$ galaxy in a typical
galaxy observed at $z \sim 0.2$. The number of minor mergers
needed to transform the $z \sim 0.6$ galaxy in a bigger and more
DM-dominated galaxy is found if the DM mass is accreting with a
lower rate and only for the AC+NFW mass profile. This is possible
if the main progenitor is merging with lower mass galaxies with
smaller \TtoSM. This result is consistent with the indications provided in terms of DM profile by the steep DM density slope found and discussed in \Sec\ref{sec:av_DM}.

{\it This result suggests that our massive galaxies at $z \sim 0.6$ have to
merge with a population of less massive galaxies with lower
total-to-stellar mass ratios.} Abundance matching studies predict that galaxies with $\mst < 10^{11.4} \, \Msun$ have, on average,
smaller \TtoSM\ (see e.g. \citealt{Moster+10}).

\section{A comment on alternatives to cold dark matter}\label{sec:alternatives}

In this paper we have presented a dynamical analysis that allows us to derive constraints on the total mass in the central regions of ETGs from simple aperture velocity dispersion measurements (e.g. from fiber or multi-slit spectroscopy). As starting working hypothesis we have assumed a strict Cold Dark Matter (CDM) paradigm (e.g. galaxies are hosted in cuspy NFW haloes with or without baryonic effects like adiabatic contraction, see Sect. \ref{sec:av_DM}). In this scenario, our results point to a significant DM fraction within \Re, which strongly depends on the assumptions made on the IMF. However, both the lack of direct evidence of the DM particle constituent and the presence of mismatches between predictions from the CDM and observations (e.g. the ``missing satellite'' problem, \citealt{Simon+07}, or the ``too big to fail'' problem, \citealt{Boylan-Kolchin+11_Too_big_to_fail}) have stimulated a wide range of alternative explanations, including different DM ``flavors'' (e.g. warm, self-interacting or fuzzy DM: \citealt{Viel+13,Tulin_Yu18_SIDM,Hu+2000_FuzzyCDM}) and even fully alternative to the DM (e.g. MOND: \citealt{Milgrom+01_MOND_review,Tortora+14_MOND}, Emergent Gravity: \citealt{Verlinde16}, no-DM $f(R)$ theories: \citealt{Napolitano+12_fR}).

Neither scenario, though, has found full observational or theoretical support.
There are simulations adopting different DM flavours, or more complex cosmologies based on dark energy models or other alternative gravity theories. However, to our knowledge, these are not providing predictions for central DM fractions in massive ellipticals we can test against the data (\citealt{Vogelsberger+20_simulations}).

From the modeling point of view, there have been a few attempts to incorporate in dynamical analyses of ETGs, some of these alternative theories.
E.g., we have demonstrated that the outcomes of particular $f(R)$ theories\footnote{These theories come the from generalization of the scalar invariant in the General Relativity.}, predicting an effective total potential including a Yukawa modification to the standard Newtonian potential, fit quite well the extended kinematic data of ellipticals (\citealt{Napolitano+12_fR}). Moreover, in \cite{Tortora+14_MOND} and \cite{Tortora+18_Verlinde}, we have shown that the central velocity dispersion of massive ETGs, can be interpreted within such alternative scenarios to DM, if a non-universal IMF is considered. Interestingly, within these alternative theories, we find taht a  bottom-heavy (bottom-light) IMF is needed in the ETGs with the largest (smallest) $\sigma$, similarly to what is found within a standard $\Lambda$CDM paradigm.

\section{Conclusions and Future perspectives}\label{sec:conclusions}
In this paper we have summarised a simple but accurate dynamical method based on Jeans equations (see \Sec\ref{subsec:Jeans}), to estimate the DM content within 1 effective radius in Early-Type Galaxies, making use of high quality (i.e. good seeing and high resolution) surface photometry and aperture internal kinematics (i.e. velocity dispersion). The method has been proved to reproduce results consistent with more accurate techniques (e.g. Schwarzschild's orbital superposition, \citealt{ThomasJ+11}; \citealt{Tortora+18_KiDS_DMevol}), if limited to the central effective radii, measuring DM fractions across cosmic time and constraining physical processes driving this evolution.

We have focused, in particular, on the study of the \TtoSM\ and \fdm\ for massive ETGs ($\log \mst/M_\odot>11.2$, Chabrier IMF) as a function of redshift, discussing the impact
of galaxy models and IMF. We base our conclusions on the results obtained with two reference samples of ETGs: one made of $z \sim 0$ local galaxies from the SPIDER project (\citealt{SPIDER-VI}) and one assembled cross-matching KiDS photometry with SDSS and BOSS spectroscopy (\citealt{Tortora+18_KiDS_DMevol}). We list below the main results.
\begin{itemize}
\item The DM fraction strongly correlates with different galaxy
parameters, e.g. the effective radius, the total mass within the effective radius, the S\'ersic index, the mean stellar density, while it has a milder correlation with stellar mass and velocity dispersion, if the IMF is taken constant (see \Sec\ref{sec:DM_correlations}).  In particular we have found that more massive and larger galaxies have a larger amount of DM. However, looking at the \TtoSM\ vs. \mst\ relation, a significant part of the scatter of the \TtoSM\ vs. \mst\ relation comes from the variation of the \TtoSM\ with redshift at fixed mass (see below), hence containing crucial information on the galaxy assembly processes. These DM fractions
are quite consistent with independent literature, as combined dynamical+lensing analysis (\citealt{Auger+10_SLACSX};
\citealt{Shajib+21}) or simulations (e.g.
\citealt{Lovell+18_Illustris}; \citealt{WangY+20_IllustrisTNG}).
\item Different lines of evidences suggest that IMF is not universal and change with galaxy parameters. \TtoSM\ analysis presented in this paper offers a simple way to reach this conclusion since higher stellar \ML\
normalizations, as those produced by a Salpter IMF, produce unphysical negative \fdm\ for those systems which tend to have smaller sizes and masses (see also \citealt{TRN13_SPIDER_IMF} and
\citealt{Tortora+14_DMevol}). If from one side the large
variation of the IMF with velocity dispersion make DM fraction
constant with this parameter, the milder correlation with sizes,
produce a strong variation of DM fraction with size.
\item The central mean DM density show indications of cuspy
DM haloes in both local and higher-redshift galaxies. Moreover,
we have confirmed the evidences of a DM non-homology first discussed in \cite{Tortora+09}, showing that observations prefer nearly isothermal total mass profiles at large masses and sizes and constant-\ML\ profiles at lower masses.
{\it This is consistent with a non-universality of the total mass density slope with mass}. In fact, lower-mass galaxies, down to $\mst \sim 3 \times 10^{10}\, \rm \Msun$, present steeper profiles, and this can be explained by a varying role of
dissipation and merging in terms of mass (\citealt{Dutton_Treu14}; \citealt{Tortora+14_DMslope,Tortora+19_LTGs_DM_and_slopes}).
\item DM fractions of massive ETGs decrease with redshift: at a given mass galaxies at higher redshift tend to have smaller \TtoSM\ than galaxies at lower redshift. On the contrary, because of the smaller \Re\ at larger redshift, the average DM density is larger at higher redshift. We have demostrated that this is an effect of minor merging of progenitor galaxies with lower-mass galaxies characterized by a smaller total-to-stellar mass ratio. Indeed, minor merging, unlike major merging, produce the correct growth of the galaxies moving from the highest redshift bin to the lower-one and still supply the necessary stellar mass for the stellar mass growth, but possibly a faster growth of the DM, which, combined with larger sizes, contributes to a faster growth of the overall \TtoSM.
\end{itemize}

The results discussed are very promising in view of the plethora of new observations that will increase data for massive ETGs up to $z=1$ and above. The simple Jeans modelling described here has the great advantage to be applicable to large galaxy samples with elementary measurements, that will be soon under the reach of all sky surveys either from ground (photometry: e.g. Vera Rubin observatory/LSST, \citealt{Izevic+19_LSST}; spectroscopy: StePS@WEAVE, \citealt{Costantin+19_steps}, 4MOST, \citealt{deJongR+11_4MOST}, DESI, \citealt{2016arXiv161100036D}) or from Space (photometry: Euclid, \citealt{Laureijs+11_Euclid}; CSST, \citealt{2018cosp...42E3821Z}).
The disadvantage is that the methods cannot constrain the stellar orbital anisotropy (which is assumed in the analysis to be reasonably constantly equal to zero in the very galaxy centers, \citealt{Tortora+09}) and is limited to the inference of the total mass; thus, the DM halo properties are model dependent (e.g. \citealt{Tortora+14_DMslope}). This makes the method equivalent to ``gravitational lensing'' only analyses (see e.g. \citealt{Auger+10_SLACSX}). Hence, the two methods are often used in combination to break mutual degeneracies and/or derive constraints on the DM density slope (e.g., \citealt{Shajib+21}). However, if one is interested to collect information on the DM content of very large samples, eventually up to millions of galaxies in the era of next generation spectroscopic surveys, we have demonstrated in this paper that the Jeans method has a great potential in providing DM estimates that can be used to test models for dark and stellar mass assembly (see e.g. \Sec\ref{subsec:evol_trends}), with lower computation times with respect to more complex approaches. Furthermore, it can be easily adapted to test alternative theories of gravity or different DM flavor predictions.

However, on smaller samples, there are two different ways to obtain more precise constraints on the DM distribution. One consists in the combination of the Jeans analysis with strong gravitational lensing and stellar population. This will be possible with new samples of strong lenses that will be discovered in future wide-field observations (e.g. from Euclid and Rubin Observatory). In particular, Euclid is expected to find $\sim 170,000$ potential galaxy-scale gravitational lenses within the $15,000$ \sqd\ of the survey (\citealt{}; \citealt{Petrillo+19_LinKS}). For a large part of these systems a measure of spectroscopic redshifts and velocity dispersions will be available, providing us and exceptional dataset for the determination of DM fractions. A second one is offered by repeated observations on individual galaxies observations from multiple instruments/surveys (see e.g. BOSS and LAMOST: \citealt{Napolitano+20_LAMOST}).
In particular, the possibility of relying on different fiber/slit apertures will provide multiple constraints on the central mass profile of these galaxies, that can be eventually further improved if combined with lensing and/or high-spatial resolution from adaptive optics (e.g. MAVIS@VLT, \citealt{McDermid+20_MAVIS}).

Finally, the brightest possible future to push forward this kind of analysis will consist of deep spectroscopic surveys measuring velocity dispersions for lower-mass and higher-z ETGs, complementing what is only possible now at very high masses ($\mst \gsim 10^{11}\, \rm \Msun$) and $z \lsim 0.7$ (e.g. \citealt{Tortora+18_KiDS_DMevol}). Collecting new observations for ETGs with $\mst \sim 3 \times 10^{10}\, \rm \Msun$ will allow us to map, as a function of cosmic time, the dynamical properties and the DM content of those galaxies which are characterized by the largest baryonic content and star formation efficiency. This ``bimodality'' mass scale, corresponding to a virial mass of $\sim 10^{12}\, \rm \Msun$, emerges from different kinds of observations and typically separate passive from star-forming systems, separating two mass regimes where galaxies are driven by different kinds of physical processes (e.g.  \citealt{Napolitano+05}; \citealt{dek_birn06,Dekel+19}; \citealt{Tortora+19_LTGs_DM_and_slopes} and references therein). Going further below this characteristic mass scale, these observations, combined with current and future cosmological simulations, will allow us to describe the DM assembly of galaxies across a wide range of masses and redshifts within a single coherent galaxy formation framework.

\section*{Acknowledgement}
We thank the two referees for the comments provided.

\bibliographystyle{frontiersinSCNS_ENG_HUMS} 



\begin{thebibliography}{133}
\providecommand{\natexlab}[1]{#1} \expandafter\ifx\csname
urlstyle\endcsname\relax
  \providecommand{\doi}[1]{doi:\discretionary{}{}{}#1}\else
  \providecommand{\doi}{doi:\discretionary{}{}{}\begingroup
  \urlstyle{rm}\Url}\fi
\providecommand{\selectlanguage}[1]{\relax}
\providecommand{\bibAnnoteFile}[1]{%
  \IfFileExists{#1}{\begin{quotation}\noindent\textsc{Key:} #1\\
  \textsc{Annotation:}\ \input{#1}\end{quotation}}{}}
\providecommand{\bibAnnote}[2]{%
  \begin{quotation}\noindent\textsc{Key:} #1\\
  \textsc{Annotation:}\ #2\end{quotation}}

\bibitem[{{Abazajian} et~al.(2009){Abazajian}, {Adelman-McCarthy},
  {Ag{\"u}eros}, {Allam}, {Allende Prieto}, {An} et~al.}]{SDSS_DR7_Abazajian}
{Abazajian}, K.~N., {Adelman-McCarthy}, J.~K., {Ag{\"u}eros},
M.~A., {Allam},
  S.~S., {Allende Prieto}, C., {An}, D., et~al. (2009).
\newblock {The Seventh Data Release of the Sloan Digital Sky Survey}.
\newblock \emph{\apjs} 182, 543.
\newblock \doi{10.1088/0067-0049/182/2/543}
\input{SDSS_DR7_Abazajian}

\bibitem[{{Alabi} et~al.(2016){Alabi}, {Forbes}, {Romanowsky}, {Brodie},
  {Strader}, {Janz} et~al.}]{Alabi+16}
{Alabi}, A.~B., {Forbes}, D.~A., {Romanowsky}, A.~J., {Brodie},
J.~P.,
  {Strader}, J., {Janz}, J., et~al. (2016).
\newblock {The SLUGGS survey: the mass distribution in early-type galaxies
  within five effective radii and beyond}.
\newblock \emph{\mnras} 460, 3838--3860.
\newblock \doi{10.1093/mnras/stw1213}
\input{Alabi+16}

\bibitem[{{Auger} et~al.(2009){Auger}, {Treu}, {Bolton}, {Gavazzi}, {Koopmans},
  {Marshall} et~al.}]{Auger+09_SLACSIX}
{Auger}, M.~W., {Treu}, T., {Bolton}, A.~S., {Gavazzi}, R.,
{Koopmans},
  L.~V.~E., {Marshall}, P.~J., et~al. (2009).
\newblock {The Sloan Lens ACS Survey. IX. Colors, Lensing, and Stellar Masses
  of Early-Type Galaxies}.
\newblock \emph{\apj} 705, 1099--1115.
\newblock \doi{10.1088/0004-637X/705/2/1099}
\input{Auger+09_SLACSIX}

\bibitem[{{Auger} et~al.(2010{\natexlab{a}}){Auger}, {Treu}, {Bolton},
  {Gavazzi}, {Koopmans}, {Marshall} et~al.}]{Auger+10_SLACSX}
{Auger}, M.~W., {Treu}, T., {Bolton}, A.~S., {Gavazzi}, R.,
{Koopmans}, L.~V.~E., {Marshall}, P.~J., et~al.
(2010{\natexlab{a}}).
\newblock {The Sloan Lens ACS Survey. X. Stellar, Dynamical, and Total Mass
  Correlations of Massive Early-type Galaxies}.
\newblock \emph{\apj} 724, 511--525.
\newblock \doi{10.1088/0004-637X/724/1/511}
\input{Auger+10_SLACSX}

\bibitem[{{Auger} et~al.(2010{\natexlab{b}}){Auger}, {Treu}, {Gavazzi},
  {Bolton}, {Koopmans}, and {Marshall}}]{Auger+10}
{Auger}, M.~W., {Treu}, T., {Gavazzi}, R., {Bolton}, A.~S.,
{Koopmans},
  L.~V.~E., and {Marshall}, P.~J. (2010{\natexlab{b}}).
\newblock {Dark Matter Contraction and the Stellar Content of Massive
  Early-type Galaxies: Disfavoring ''Light'' Initial Mass Functions}.
\newblock \emph{\apjl} 721, L163--L167.
\newblock \doi{10.1088/2041-8205/721/2/L163}
\input{Auger+10}

\bibitem[{{Beifiori} et~al.(2014){Beifiori}, {Thomas}, {Maraston}, {Steele},
  {Masters}, {Pforr} et~al.}]{Beifiori+14}
{Beifiori}, A., {Thomas}, D., {Maraston}, C., {Steele}, O.,
{Masters}, K.~L.,
  {Pforr}, J., et~al. (2014).
\newblock {Redshift Evolution of the Dynamical Properties of Massive Galaxies
  from SDSS-III/BOSS}.
\newblock \emph{\apj} 789, 92.
\newblock \doi{10.1088/0004-637X/789/2/92}
\input{Beifiori+14}

\bibitem[{{Benson} et~al.(2000){Benson}, {Cole}, {Frenk}, {Baugh}, and
  {Lacey}}]{Benson+00}
{Benson}, A.~J., {Cole}, S., {Frenk}, C.~S., {Baugh}, C.~M., and
{Lacey}, C.~G.
  (2000).
\newblock {The nature of galaxy bias and clustering}.
\newblock \emph{\mnras} 311, 793--808.
\newblock \doi{10.1046/j.1365-8711.2000.03101.x}
\input{Benson+00}

\bibitem[{{Boylan-Kolchin} et~al.(2011){Boylan-Kolchin}, {Bullock}, and
  {Kaplinghat}}]{Boylan-Kolchin+11_Too_big_to_fail}
{Boylan-Kolchin}, M., {Bullock}, J.~S., and {Kaplinghat}, M.
(2011).
\newblock {Too big to fail? The puzzling darkness of massive Milky Way
  subhaloes}.
\newblock \emph{\mnras} 415, L40--L44.
\newblock \doi{10.1111/j.1745-3933.2011.01074.x}
\input{Boylan-Kolchin+11_Too_big_to_fail}

\bibitem[{{Boylan-Kolchin} et~al.(2005){Boylan-Kolchin}, {Ma}, and
  {Quataert}}]{Boylan-Kolchin+05}
{Boylan-Kolchin}, M., {Ma}, C.-P., and {Quataert}, E. (2005).
\newblock {Dissipationless mergers of elliptical galaxies and the evolution of
  the fundamental plane}.
\newblock \emph{\mnras} 362, 184--196.
\newblock \doi{10.1111/j.1365-2966.2005.09278.x}
\input{Boylan-Kolchin+05}

\bibitem[{{Buitrago} et~al.(2008){Buitrago}, {Trujillo}, {Conselice},
  {Bouwens}, {Dickinson}, and {Yan}}]{Buitrago+08}
{Buitrago}, F., {Trujillo}, I., {Conselice}, C.~J., {Bouwens},
R.~J.,
  {Dickinson}, M., and {Yan}, H. (2008).
\newblock {Size Evolution of the Most Massive Galaxies at 1.7 $<$ z $<$ 3
  from GOODS NICMOS Survey Imaging}.
\newblock \emph{\apjl} 687, L61.
\newblock \doi{10.1086/592836}
\input{Buitrago+08}

\bibitem[{{Burkert}(1995)}]{Burkert95}
{Burkert}, A. (1995).
\newblock {The Structure of Dark Matter Halos in Dwarf Galaxies}.
\newblock \emph{\apjl} 447, L25.
\newblock \doi{10.1086/309560}
\input{Burkert95}

\bibitem[{{Cappellari} et~al.(2006){Cappellari}, {Bacon}, {Bureau}, {Damen},
  {Davies}, {de Zeeuw} et~al.}]{Cappellari+06}
{Cappellari}, M., {Bacon}, R., {Bureau}, M., {Damen}, M.~C.,
{Davies}, R.~L.,
  {de Zeeuw}, P.~T., et~al. (2006).
\newblock {The SAURON project - IV. The mass-to-light ratio, the virial mass
  estimator and the Fundamental Plane of elliptical and lenticular galaxies}.
\newblock \emph{\mnras} 366, 1126--1150.
\newblock \doi{10.1111/j.1365-2966.2005.09981.x}
\input{Cappellari+06}

\bibitem[{{Cappellari} et~al.(2012){Cappellari}, {McDermid}, {Alatalo},
  {Blitz}, {Bois}, {Bournaud} et~al.}]{Cappellari+12}
{Cappellari}, M., {McDermid}, R.~M., {Alatalo}, K., {Blitz}, L.,
{Bois}, M.,
  {Bournaud}, F., et~al. (2012).
\newblock {Systematic variation of the stellar initial mass function in
  early-type galaxies}.
\newblock \emph{\nat} 484, 485--488.
\newblock \doi{10.1038/nature10972}
\input{Cappellari+12}

\bibitem[{{Cappellari} et~al.(2013{\natexlab{a}}){Cappellari}, {McDermid},
  {Alatalo}, {Blitz}, {Bois}, {Bournaud} et~al.}]{Cappellari+13_ATLAS3D_XX}
{Cappellari}, M., {McDermid}, R.~M., {Alatalo}, K., {Blitz}, L.,
{Bois}, M.,
  {Bournaud}, F., et~al. (2013{\natexlab{a}}).
\newblock {The ATLAS$^{3D}$ project - XX. Mass-size and mass-{$\sigma$}
  distributions of early-type galaxies: bulge fraction drives kinematics,
  mass-to-light ratio, molecular gas fraction and stellar initial mass
  function}.
\newblock \emph{\mnras} 432, 1862--1893.
\newblock \doi{10.1093/mnras/stt644}
\input{Cappellari+13_ATLAS3D_XX}

\bibitem[{{Cappellari} et~al.(2013{\natexlab{b}}){Cappellari}, {Scott},
  {Alatalo}, {Blitz}, {Bois}, {Bournaud} et~al.}]{Cappellari+13_ATLAS3D_XV}
{Cappellari}, M., {Scott}, N., {Alatalo}, K., {Blitz}, L., {Bois},
M.,
  {Bournaud}, F., et~al. (2013{\natexlab{b}}).
\newblock {The ATLAS$^{3D}$ project - XV. Benchmark for early-type galaxies
  scaling relations from 260 dynamical models: mass-to-light ratio, dark
  matter, Fundamental Plane and Mass Plane}.
\newblock \emph{\mnras} 432, 1709--1741.
\newblock \doi{10.1093/mnras/stt562}
\input{Cappellari+13_ATLAS3D_XV}

\bibitem[{{Cardone} et~al.(2011){Cardone}, {Del Popolo}, {Tortora}, and
  {Napolitano}}]{Cardone+11SIM}
{Cardone}, V.~F., {Del Popolo}, A., {Tortora}, C., and
{Napolitano}, N.~R.
  (2011).
\newblock {Secondary infall model and dark matter scaling relations in
  intermediate-redshift early-type galaxies}.
\newblock \emph{\mnras} 416, 1822--1835.
\newblock \doi{10.1111/j.1365-2966.2011.19162.x}
\input{Cardone+11SIM}

\bibitem[{{Cardone} and {Tortora}(2010)}]{CT10}
{Cardone}, V.~F. and {Tortora}, C. (2010).
\newblock {Dark matter scaling relations in intermediate z haloes}.
\newblock \emph{\mnras} 409, 1570--1576.
\newblock \doi{10.1111/j.1365-2966.2010.17398.x}
\input{CT10}

\bibitem[{{Cardone} et~al.(2009){Cardone}, {Tortora}, {Molinaro}, and
  {Salzano}}]{Cardone+09}
{Cardone}, V.~F., {Tortora}, C., {Molinaro}, R., and {Salzano}, V.
(2009).
\newblock {The global mass-to-light ratio of SLACS lenses}.
\newblock \emph{\aap} 504, 769--788.
\newblock \doi{10.1051/0004-6361/200811090}
\input{Cardone+09}

\bibitem[{{Cenarro} and {Trujillo}(2009)}]{Cenarro_Trujillo09}
{Cenarro}, A.~J. and {Trujillo}, I. (2009).
\newblock {Mild Velocity Dispersion Evolution of Spheroid-Like Massive Galaxies
  Since z \~{} 2}.
\newblock \emph{\apjl} 696, L43--L47.
\newblock \doi{10.1088/0004-637X/696/1/L43}
\input{Cenarro_Trujillo09}

\bibitem[{{Chabrier}(2001)}]{Chabrier01}
{Chabrier}, G. (2001).
\newblock {The Galactic Disk Mass Budget. I. Stellar Mass Function and
  Density}.
\newblock \emph{\apj} 554, 1274--1281.
\newblock \doi{10.1086/321401}
\input{Chabrier01}

\bibitem[{{Chabrier}(2003)}]{Chabrier03}
{Chabrier}, G. (2003).
\newblock {Galactic Stellar and Substellar Initial Mass Function}.
\newblock \emph{\pasp} 115, 763--795.
\newblock \doi{10.1086/376392}
\input{Chabrier03}

\bibitem[{{Conroy} and {van Dokkum}(2012)}]{Conroy_vanDokkum12b}
{Conroy}, C. and {van Dokkum}, P.~G. (2012).
\newblock {The Stellar Initial Mass Function in Early-type Galaxies From
  Absorption Line Spectroscopy. II. Results}.
\newblock \emph{\apj} 760, 71.
\newblock \doi{10.1088/0004-637X/760/1/71}
\input{Conroy_vanDokkum12b}

\bibitem[{{Conroy} and {Wechsler}(2009)}]{CW09}
{Conroy}, C. and {Wechsler}, R.~H. (2009).
\newblock {Connecting Galaxies, Halos, and Star Formation Rates Across Cosmic
  Time}.
\newblock \emph{\apj} 696, 620--635.
\newblock \doi{10.1088/0004-637X/696/1/620}
\input{CW09}

\bibitem[{{Conselice}(2014)}]{Conselice14_review}
{Conselice}, C.~J. (2014).
\newblock {The Evolution of Galaxy Structure Over Cosmic Time}.
\newblock \emph{\araa} 52, 291--337.
\newblock \doi{10.1146/annurev-astro-081913-040037}
\input{Conselice14_review}

\bibitem[{{Corsini} et~al.(2017){Corsini}, {Wegner}, {Thomas}, {Saglia}, and
  {Bender}}]{Corsini+17}
{Corsini}, E.~M., {Wegner}, G.~A., {Thomas}, J., {Saglia}, R.~P.,
and {Bender},
  R. (2017).
\newblock {The density of dark matter haloes of early-type galaxies in
  low-density environments}.
\newblock \emph{\mnras} 466, 974--995.
\newblock \doi{10.1093/mnras/stw2935}
\input{Corsini+17}

\bibitem[{{Costantin} et~al.(2019){Costantin}, {Iovino}, {Zibetti},
  {Longhetti}, {Gallazzi}, {Mercurio} et~al.}]{Costantin+19_steps}
{Costantin}, L., {Iovino}, A., {Zibetti}, S., {Longhetti}, M.,
{Gallazzi}, A.,
  {Mercurio}, A., et~al. (2019).
\newblock {A few StePS forward in unveiling the complexity of galaxy evolution:
  light-weighted stellar ages of intermediate-redshift galaxies with WEAVE}.
\newblock \emph{\aap} 632, A9.
\newblock \doi{10.1051/0004-6361/201936550}
\input{Costantin+19_steps}

\bibitem[{{Courteau} et~al.(2014){Courteau}, {Cappellari}, {de Jong}, {Dutton},
  {Emsellem}, {Hoekstra} et~al.}]{Courteau+14_review}
{Courteau}, S., {Cappellari}, M., {de Jong}, R.~S., {Dutton},
A.~A.,
  {Emsellem}, E., {Hoekstra}, H., et~al. (2014).
\newblock {Galaxy masses}.
\newblock \emph{Reviews of Modern Physics} 86, 47--119.
\newblock \doi{10.1103/RevModPhys.86.47}
\input{Courteau+14_review}

\bibitem[{{Daddi} et~al.(2005){Daddi}, {Renzini}, {Pirzkal}, {Cimatti},
  {Malhotra}, {Stiavelli} et~al.}]{Daddi+05}
{Daddi}, E., {Renzini}, A., {Pirzkal}, N., {Cimatti}, A.,
{Malhotra}, S.,
  {Stiavelli}, M., et~al. (2005).
\newblock {Passively Evolving Early-Type Galaxies at $1.4 < z < 2.5$ in the Hubble Ultra Deep Field}.
\newblock \emph{\apj} 626, 680--697.
\newblock \doi{10.1086/430104}
\input{Daddi+05}

\bibitem[{{de Jong} et~al.(2017){de Jong}, {Kleijn}, {Erben}, {Hildebrandt},
  {Kuijken}, {Sikkema} et~al.}]{deJong+17_KiDS_DR3}
{de Jong}, J.~T.~A., {Kleijn}, G.~A.~V., {Erben}, T.,
{Hildebrandt}, H.,
  {Kuijken}, K., {Sikkema}, G., et~al. (2017).
\newblock {The third data release of the Kilo-Degree Survey and associated data
  products}.
\newblock \emph{\aap} 604, A134.
\newblock \doi{10.1051/0004-6361/201730747}
\input{deJong+17_KiDS_DR3}

\bibitem[{{de Jong} et~al.(2015){de Jong}, {Verdoes Kleijn}, {Boxhoorn},
  {Buddelmeijer}, {Capaccioli}, {Getman} et~al.}]{deJong+15_KiDS_paperI}
{de Jong}, J.~T.~A., {Verdoes Kleijn}, G.~A., {Boxhoorn}, D.~R.,
  {Buddelmeijer}, H., {Capaccioli}, M., {Getman}, F., et~al. (2015).
\newblock {The first and second data releases of the Kilo-Degree Survey}.
\newblock \emph{\aap} 582, A62.
\newblock \doi{10.1051/0004-6361/201526601}
\input{deJong+15_KiDS_paperI}

\bibitem[{{de Jong}(2011)}]{deJongR+11_4MOST}
{de Jong}, R. (2011).
\newblock {4MOST {\textemdash} 4-metre Multi-Object Spectroscopic Telescope}.
\newblock \emph{The Messenger} 145, 14--16
\input{deJongR+11_4MOST}

\bibitem[{{Dekel} and {Birnboim}(2006)}]{dek_birn06}
{Dekel}, A. and {Birnboim}, Y. (2006).
\newblock {Galaxy bimodality due to cold flows and shock heating}.
\newblock \emph{\mnras} 368, 2--20.
\newblock \doi{10.1111/j.1365-2966.2006.10145.x}
\input{dek_birn06}

\bibitem[{{Dekel} and {Burkert}(2014)}]{Dekel_Burkert14}
{Dekel}, A. and {Burkert}, A. (2014).
\newblock {Wet disc contraction to galactic blue nuggets and quenching to red
  nuggets}.
\newblock \emph{\mnras} 438, 1870--1879.
\newblock \doi{10.1093/mnras/stt2331}
\input{Dekel_Burkert14}

\bibitem[{{Dekel} et~al.(2019){Dekel}, {Lapiner}, and {Dubois}}]{Dekel+19}
{Dekel}, A., {Lapiner}, S., and {Dubois}, Y. (2019).
\newblock {Origin of the Golden Mass of Galaxies and Black Holes}.
\newblock \emph{arXiv e-prints} , arXiv:1904.08431
\input{Dekel+19}

\bibitem[{{DESI Collaboration} et~al.(2016){DESI Collaboration}, {Aghamousa},
  {Aguilar}, {Ahlen}, {Alam}, {Allen} et~al.}]{2016arXiv161100036D}
{DESI Collaboration}, {Aghamousa}, A., {Aguilar}, J., {Ahlen}, S.,
{Alam}, S.,
  {Allen}, L.~E., et~al. (2016).
\newblock {The DESI Experiment Part I: Science,Targeting, and Survey Design}.
\newblock \emph{arXiv e-prints} , arXiv:1611.00036
\input{2016arXiv161100036D}

\bibitem[{{Dom{\'\i}nguez S{\'a}nchez} et~al.(2019){Dom{\'\i}nguez
  S{\'a}nchez}, {Bernardi}, {Brownstein}, {Drory}, and
  {Sheth}}]{2019MNRAS.489.5612D}
{Dom{\'\i}nguez S{\'a}nchez}, H., {Bernardi}, M., {Brownstein},
J.~R., {Drory},
  N., and {Sheth}, R.~K. (2019).
\newblock {Galaxy properties as revealed by MaNGA - I. Constraints on IMF and
  M$_{*}$/L gradients in ellipticals}.
\newblock \emph{\mnras} 489, 5612--5632.
\newblock \doi{10.1093/mnras/stz2414}
\input{2019MNRAS.489.5612D}

\bibitem[{{Dutton} and {Treu}(2014)}]{Dutton_Treu14}
{Dutton}, A.~A. and {Treu}, T. (2014).
\newblock {The bulge-halo conspiracy in massive elliptical galaxies:
  implications for the stellar initial mass function and halo response to
  baryonic processes}.
\newblock \emph{\mnras} 438, 3594--3602.
\newblock \doi{10.1093/mnras/stt2489}
\input{Dutton_Treu14}

\bibitem[{{Fan} et~al.(2008){Fan}, {Lapi}, {De Zotti}, and {Danese}}]{Fan+08}
{Fan}, L., {Lapi}, A., {De Zotti}, G., and {Danese}, L. (2008).
\newblock {The Dramatic Size Evolution of Elliptical Galaxies and the Quasar
  Feedback}.
\newblock \emph{\apjl} 689, L101--L104.
\newblock \doi{10.1086/595784}
\input{Fan+08}

\bibitem[{{Gallazzi} et~al.(2005){Gallazzi}, {Charlot}, {Brinchmann}, {White},
  and {Tremonti}}]{Gallazzi+05}
{Gallazzi}, A., {Charlot}, S., {Brinchmann}, J., {White},
S.~D.~M., and
  {Tremonti}, C.~A. (2005).
\newblock {The ages and metallicities of galaxies in the local universe}.
\newblock \emph{\mnras} 362, 41--58.
\newblock \doi{10.1111/j.1365-2966.2005.09321.x}
\input{Gallazzi+05}

\bibitem[{{Gavazzi} et~al.(2007){Gavazzi}, {Treu}, {Rhodes}, {Koopmans},
  {Bolton}, {Burles} et~al.}]{Gavazzi+07_SLACSIV}
{Gavazzi}, R., {Treu}, T., {Rhodes}, J.~D., {Koopmans}, L.~V.~E.,
{Bolton},
  A.~S., {Burles}, S., et~al. (2007).
\newblock {The Sloan Lens ACS Survey. IV. The Mass Density Profile of
  Early-Type Galaxies out to 100 Effective Radii}.
\newblock \emph{\apj} 667, 176--190.
\newblock \doi{10.1086/519237}
\input{Gavazzi+07_SLACSIV}

\bibitem[{{Gnedin} et~al.(2004){Gnedin}, {Kravtsov}, {Klypin}, and
  {Nagai}}]{Gnedin+04}
{Gnedin}, O.~Y., {Kravtsov}, A.~V., {Klypin}, A.~A., and {Nagai},
D. (2004).
\newblock {Response of Dark Matter Halos to Condensation of Baryons:
  Cosmological Simulations and Improved Adiabatic Contraction Model}.
\newblock \emph{\apj} 616, 16--26.
\newblock \doi{10.1086/424914}
\input{Gnedin+04}

\bibitem[{{Goudfrooij} and {Kruijssen}(2013)}]{Goudfrooij_Kruijssen13}
{Goudfrooij}, P. and {Kruijssen}, J.~M.~D. (2013).
\newblock {The Optical Colors of Giant Elliptical Galaxies and their Metal-Rich
  Globular Clusters Indicate a Bottom-Heavy Initial Mass Function}.
\newblock \emph{\apj} 762, 107.
\newblock \doi{10.1088/0004-637X/762/2/107}
\input{Goudfrooij_Kruijssen13}

\bibitem[{{Grillo}(2010)}]{Grillo10}
{Grillo}, C. (2010).
\newblock {Projected Central Dark Matter Fractions and Densities in Massive
  Early-type Galaxies from the Sloan Digital Sky Survey}.
\newblock \emph{\apj} 722, 779--787.
\newblock \doi{10.1088/0004-637X/722/1/779}
\input{Grillo10}

\bibitem[{{Grillo} and {Gobat}(2010)}]{Grillo_Cobat10}
{Grillo}, C. and {Gobat}, R. (2010).
\newblock {On the initial mass function and tilt of the fundamental plane of
  massive early-type galaxies}.
\newblock \emph{\mnras} 402, L67--L71.
\newblock \doi{10.1111/j.1745-3933.2009.00803.x}
\input{Grillo_Cobat10}

\bibitem[{{Grillo} et~al.(2009){Grillo}, {Gobat}, {Lombardi}, and
  {Rosati}}]{Grillo+09}
{Grillo}, C., {Gobat}, R., {Lombardi}, M., and {Rosati}, P.
(2009).
\newblock {Photometric mass and mass decomposition in early-type lens
  galaxies}.
\newblock \emph{\aap} 501, 461--474.
\newblock \doi{10.1051/0004-6361/200811604}
\input{Grillo+09}

\bibitem[{{Hilz} et~al.(2013){Hilz}, {Naab}, and {Ostriker}}]{Hilz+13}
{Hilz}, M., {Naab}, T., and {Ostriker}, J.~P. (2013).
\newblock {How do minor mergers promote inside-out growth of ellipticals,
  transforming the size, density profile and dark matter fraction?}
\newblock \emph{\mnras} 429, 2924--2933.
\newblock \doi{10.1093/mnras/sts501}
\input{Hilz+13}

\bibitem[{{Hopkins} et~al.(2010){Hopkins}, {Croton}, {Bundy}, {Khochfar}, {van
  den Bosch}, {Somerville} et~al.}]{Hopkins+10_Mergers_LCDM}
{Hopkins}, P.~F., {Croton}, D., {Bundy}, K., {Khochfar}, S., {van
den Bosch},
  F., {Somerville}, R.~S., et~al. (2010).
\newblock {Mergers in {$\Lambda$}CDM: Uncertainties in Theoretical Predictions
  and Interpretations of the Merger Rate}.
\newblock \emph{\apj} 724, 915--945.
\newblock \doi{10.1088/0004-637X/724/2/915}
\input{Hopkins+10_Mergers_LCDM}

\bibitem[{{Hopkins} et~al.(2009){Hopkins}, {Hernquist}, {Cox}, {Keres}, and
  {Wuyts}}]{Hopkins+09_DELGN_IV}
{Hopkins}, P.~F., {Hernquist}, L., {Cox}, T.~J., {Keres}, D., and
{Wuyts}, S.
  (2009).
\newblock {Dissipation and Extra Light in Galactic Nuclei. IV. Evolution in the
  Scaling Relations of Spheroids}.
\newblock \emph{\apj} 691, 1424--1458.
\newblock \doi{10.1088/0004-637X/691/2/1424}
\input{Hopkins+09_DELGN_IV}

\bibitem[{{Hu} et~al.(2000){Hu}, {Barkana}, and {Gruzinov}}]{Hu+2000_FuzzyCDM}
{Hu}, W., {Barkana}, R., and {Gruzinov}, A. (2000).
\newblock {Fuzzy Cold Dark Matter: The Wave Properties of Ultralight
  Particles}.
\newblock \emph{\prl} 85, 1158--1161.
\newblock \doi{10.1103/PhysRevLett.85.1158}
\input{Hu+2000_FuzzyCDM}

\bibitem[{{Hyde} and {Bernardi}(2009{\natexlab{a}})}]{HB09_curv}
{Hyde}, J.~B. and {Bernardi}, M. (2009{\natexlab{a}}).
\newblock {Curvature in the scaling relations of early-type galaxies}.
\newblock \emph{\mnras} 394, 1978--1990.
\newblock \doi{10.1111/j.1365-2966.2009.14445.x}
\input{HB09_curv}

\bibitem[{{Hyde} and {Bernardi}(2009{\natexlab{b}})}]{HB09_FP}
{Hyde}, J.~B. and {Bernardi}, M. (2009{\natexlab{b}}).
\newblock {The luminosity and stellar mass Fundamental Plane of early-type
  galaxies}.
\newblock \emph{\mnras} 396, 1171--1185.
\newblock \doi{10.1111/j.1365-2966.2009.14783.x}
\input{HB09_FP}

\bibitem[{{Ivezi{\'c}} et~al.(2019){Ivezi{\'c}}, {Kahn}, {Tyson}, {Abel},
  {Acosta}, {Allsman} et~al.}]{Izevic+19_LSST}
{Ivezi{\'c}}, {\v{Z}}., {Kahn}, S.~M., {Tyson}, J.~A., {Abel}, B.,
{Acosta},
  E., {Allsman}, R., et~al. (2019).
\newblock {LSST: From Science Drivers to Reference Design and Anticipated Data
  Products}.
\newblock \emph{\apj} 873, 111.
\newblock \doi{10.3847/1538-4357/ab042c}
\input{Izevic+19_LSST}

\bibitem[{{Kauffmann} et~al.(2003){Kauffmann}, {Heckman}, {White}, {Charlot},
  {Tremonti}, {Peng} et~al.}]{Kauffmann+03}
{Kauffmann}, G., {Heckman}, T.~M., {White}, S.~D.~M., {Charlot},
S.,
  {Tremonti}, C., {Peng}, E.~W., et~al. (2003).
\newblock {The dependence of star formation history and internal structure on
  stellar mass for 10$^{5}$ low-redshift galaxies}.
\newblock \emph{\mnras} 341, 54--69.
\newblock \doi{10.1046/j.1365-8711.2003.06292.x}
\input{Kauffmann+03}

\bibitem[{{Koopmans} et~al.(2009){Koopmans}, {Bolton}, {Treu}, {Czoske},
  {Auger}, {Barnab{\`e}} et~al.}]{Koopmans+09}
{Koopmans}, L.~V.~E., {Bolton}, A., {Treu}, T., {Czoske}, O.,
{Auger}, M.~W.,
  {Barnab{\`e}}, M., et~al. (2009).
\newblock {The Structure and Dynamics of Massive Early-Type Galaxies: On
  Homology, Isothermality, and Isotropy Inside One Effective Radius}.
\newblock \emph{\apjl} 703, L51--L54.
\newblock \doi{10.1088/0004-637X/703/1/L51}
\input{Koopmans+09}

\bibitem[{{Koopmans} et~al.(2006){Koopmans}, {Treu}, {Bolton}, {Burles}, and
  {Moustakas}}]{Koopmans+06_SLACSIII}
{Koopmans}, L.~V.~E., {Treu}, T., {Bolton}, A.~S., {Burles}, S.,
and
  {Moustakas}, L.~A. (2006).
\newblock {The Sloan Lens ACS Survey. III. The Structure and Formation of
  Early-Type Galaxies and Their Evolution since z \~{} 1}.
\newblock \emph{\apj} 649, 599--615.
\newblock \doi{10.1086/505696}
\input{Koopmans+06_SLACSIII}

\bibitem[{{Kroupa}(2001)}]{Kroupa01}
{Kroupa}, P. (2001).
\newblock {On the variation of the initial mass function}.
\newblock \emph{\mnras} 322, 231--246.
\newblock \doi{10.1046/j.1365-8711.2001.04022.x}
\input{Kroupa01}

\bibitem[{{La Barbera} et~al.(2010){La Barbera}, {de Carvalho}, {de La Rosa},
  {Lopes}, {Kohl-Moreira}, and {Capelato}}]{SPIDER-I}
{La Barbera}, F., {de Carvalho}, R.~R., {de La Rosa}, I.~G.,
{Lopes}, P.~A.~A.,
  {Kohl-Moreira}, J.~L., and {Capelato}, H.~V. (2010).
\newblock {SPIDER - I. Sample and galaxy parameters in the grizYJHK wavebands}.
\newblock \emph{\mnras} 408, 1313--1334.
\newblock \doi{10.1111/j.1365-2966.2010.16850.x}
\input{SPIDER-I}

\bibitem[{{La Barbera} et~al.(2008){La Barbera}, {de Carvalho}, {Kohl-Moreira},
  {Gal}, {Soares-Santos}, {Capaccioli} et~al.}]{LaBarbera_08_2DPHOT}
{La Barbera}, F., {de Carvalho}, R.~R., {Kohl-Moreira}, J.~L.,
{Gal}, R.~R.,
  {Soares-Santos}, M., {Capaccioli}, M., et~al. (2008).
\newblock {2DPHOT: A Multi-Purpose Environment for the Two-Dimensional Analysis
  of Wide-Field Images}.
\newblock \emph{\pasp} 120, 681--702.
\newblock \doi{10.1086/588614}
\input{LaBarbera_08_2DPHOT}

\bibitem[{{La Barbera} et~al.(2013){La Barbera}, {Ferreras}, {Vazdekis}, {de la
  Rosa}, {de Carvalho}, {Trevisan} et~al.}]{LaBarbera+13_SPIDERVIII_IMF}
{La Barbera}, F., {Ferreras}, I., {Vazdekis}, A., {de la Rosa},
I.~G., {de
  Carvalho}, R.~R., {Trevisan}, M., et~al. (2013).
\newblock {SPIDER VIII - constraints on the stellar initial mass function of
  early-type galaxies from a variety of spectral features}.
\newblock \emph{\mnras} 433, 3017--3047.
\newblock \doi{10.1093/mnras/stt943}
\input{LaBarbera+13_SPIDERVIII_IMF}

\bibitem[{{Laureijs} et~al.(2011){Laureijs}, {Amiaux}, {Arduini},
  {Augu{\`e}res}, {Brinchmann}, {Cole} et~al.}]{Laureijs+11_Euclid}
{Laureijs}, R., {Amiaux}, J., {Arduini}, S., {Augu{\`e}res},
J.~L.,
  {Brinchmann}, J., {Cole}, R., et~al. (2011).
\newblock {Euclid Definition Study Report}.
\newblock \emph{arXiv e-prints} , arXiv:1110.3193
\input{Laureijs+11_Euclid}

\bibitem[{{Li} et~al.(2017){Li}, {Ge}, {Mao}, {Cappellari}, {Long}, {Li}
  et~al.}]{Li+17_IMF}
{Li}, H., {Ge}, J., {Mao}, S., {Cappellari}, M., {Long}, R.~J.,
{Li}, R.,
  et~al. (2017).
\newblock {SDSS-IV MaNGA: Variation of the Stellar Initial Mass Function in
  Spiral and Early-type Galaxies}.
\newblock \emph{\apj} 838, 77.
\newblock \doi{10.3847/1538-4357/aa662a}
\input{Li+17_IMF}

\bibitem[{{Liske} et~al.(2015){Liske}, {Baldry}, {Driver}, {Tuffs}, {Alpaslan},
  {Andrae} et~al.}]{Liske+15_GAMA}
{Liske}, J., {Baldry}, I.~K., {Driver}, S.~P., {Tuffs}, R.~J.,
{Alpaslan}, M.,
  {Andrae}, E., et~al. (2015).
\newblock {Galaxy And Mass Assembly (GAMA): end of survey report and data
  release 2}.
\newblock \emph{\mnras} 452, 2087--2126.
\newblock \doi{10.1093/mnras/stv1436}
\input{Liske+15_GAMA}

\bibitem[{{Lovell} et~al.(2018){Lovell}, {Pillepich}, {Genel}, {Nelson},
  {Springel}, {Pakmor} et~al.}]{Lovell+18_Illustris}
{Lovell}, M.~R., {Pillepich}, A., {Genel}, S., {Nelson}, D.,
{Springel}, V.,
  {Pakmor}, R., et~al. (2018).
\newblock {The fraction of dark matter within galaxies from the IllustrisTNG
  simulations}.
\newblock \emph{\mnras} \doi{10.1093/mnras/sty2339}
\input{Lovell+18_Illustris}

\bibitem[{{Macci{\`o}} et~al.(2008){Macci{\`o}}, {Dutton}, and {van den
  Bosch}}]{Maccio+08}
{Macci{\`o}}, A.~V., {Dutton}, A.~A., and {van den Bosch}, F.~C.
(2008).
\newblock {Concentration, spin and shape of dark matter haloes as a function of
  the cosmological model: WMAP1, WMAP3 and WMAP5 results}.
\newblock \emph{\mnras} 391, 1940--1954.
\newblock \doi{10.1111/j.1365-2966.2008.14029.x}
\input{Maccio+08}

\bibitem[{{Mamon} and {{\L}okas}(2005)}]{ML05a}
{Mamon}, G.~A. and {{\L}okas}, E.~L. (2005).
\newblock {Dark matter in elliptical galaxies - I. Is the total mass density
  profile of the NFW form or even steeper?}
\newblock \emph{\mnras} 362, 95--109.
\newblock \doi{10.1111/j.1365-2966.2005.09225.x}
\input{ML05a}

\bibitem[{{Mamon} and {{\L}okas}(2006)}]{ML06_erratum}
{Mamon}, G.~A. and {{\L}okas}, E.~L. (2006).
\newblock {Erratum: Dark matter in elliptical galaxies - I. Is the total mass
  density profile of the NFW form or even steeper?}
\newblock \emph{\mnras} 370, 1581--1581.
\newblock \doi{10.1111/j.1365-2966.2006.10647.x}
\input{ML06_erratum}

\bibitem[{{Mandelbaum} et~al.(2006){Mandelbaum}, {Seljak}, {Kauffmann},
  {Hirata}, and {Brinkmann}}]{Mandelbaum+06}
{Mandelbaum}, R., {Seljak}, U., {Kauffmann}, G., {Hirata}, C.~M.,
and
  {Brinkmann}, J. (2006).
\newblock {Galaxy halo masses and satellite fractions from galaxy-galaxy
  lensing in the Sloan Digital Sky Survey: stellar mass, luminosity, morphology
  and environment dependencies}.
\newblock \emph{\mnras} 368, 715--731.
\newblock \doi{10.1111/j.1365-2966.2006.10156.x}
\input{Mandelbaum+06}

\bibitem[{{Maraston} et~al.(2013){Maraston}, {Pforr}, {Henriques}, {Thomas},
  {Wake}, {Brownstein} et~al.}]{Maraston+13_BOSS}
{Maraston}, C., {Pforr}, J., {Henriques}, B.~M., {Thomas}, D.,
{Wake}, D.,
  {Brownstein}, J.~R., et~al. (2013).
\newblock {Stellar masses of SDSS-III/BOSS galaxies at z $\sim$ 0.5 and
  constraints to galaxy formation models}.
\newblock \emph{\mnras} 435, 2764--2792.
\newblock \doi{10.1093/mnras/stt1424}
\input{Maraston+13_BOSS}

\bibitem[{{Marinoni} and {Hudson}(2002)}]{MH02}
{Marinoni}, C. and {Hudson}, M.~J. (2002).
\newblock {The Mass-to-Light Function of Virialized Systems and the
  Relationship between Their Optical and X-Ray Properties}.
\newblock \emph{\apj} 569, 101--111.
\newblock \doi{10.1086/339319}
\input{MH02}

\bibitem[{{Mart\'in-Navarro} et~al.(2015){Mart\'in-Navarro}, {Barbera},
  {Vazdekis}, {Falc{\'o}n-Barroso}, and
  {Ferreras}}]{Martin-Navarro+15_IMF_variation}
{Mart\'in-Navarro}, I., {Barbera}, F.~L., {Vazdekis}, A.,
{Falc{\'o}n-Barroso},
  J., and {Ferreras}, I. (2015).
\newblock {Radial variations in the stellar initial mass function of early-type
  galaxies}.
\newblock \emph{\mnras} 447, 1033--1048.
\newblock \doi{10.1093/mnras/stu2480}
\input{Martin-Navarro+15_IMF_variation}

\bibitem[{{McDermid} et~al.(2014){McDermid}, {Cappellari}, {Alatalo}, {Bayet},
  {Blitz}, {Bois} et~al.}]{McDermid+14_IMF}
{McDermid}, R.~M., {Cappellari}, M., {Alatalo}, K., {Bayet}, E.,
{Blitz}, L.,
  {Bois}, M., et~al. (2014).
\newblock {Connection between Dynamically Derived Initial Mass Function
  Normalization and Stellar Population Parameters}.
\newblock \emph{\apjl} 792, L37.
\newblock \doi{10.1088/2041-8205/792/2/L37}
\input{McDermid+14_IMF}

\bibitem[{{McDermid} et~al.(2020){McDermid}, {Cresci}, {Rigaut}, {Bouret}, {De
  Silva}, {Gullieuszik} et~al.}]{McDermid+20_MAVIS}
{McDermid}, R.~M., {Cresci}, G., {Rigaut}, F., {Bouret}, J.-C.,
{De Silva}, G.,
  {Gullieuszik}, M., et~al. (2020).
\newblock {Phase A Science Case for MAVIS -- The Multi-conjugate
  Adaptive-optics Visible Imager-Spectrograph for the VLT Adaptive Optics
  Facility}.
\newblock \emph{arXiv e-prints} , arXiv:2009.09242
\input{McDermid+20_MAVIS}

\bibitem[{{Milgrom}(2001)}]{Milgrom+01_MOND_review}
{Milgrom}, M. (2001).
\newblock {MOND --- a Pedagogical Review}.
\newblock \emph{Acta Physica Polonica B} 32, 3613
\input{Milgrom+01_MOND_review}

\bibitem[{{Moster} et~al.(2010){Moster}, {Somerville}, {Maulbetsch}, {van den
  Bosch}, {Macci{\`o}}, {Naab} et~al.}]{Moster+10}
{Moster}, B.~P., {Somerville}, R.~S., {Maulbetsch}, C., {van den
Bosch}, F.~C.,
  {Macci{\`o}}, A.~V., {Naab}, T., et~al. (2010).
\newblock {Constraints on the Relationship between Stellar Mass and Halo Mass
  at Low and High Redshift}.
\newblock \emph{\apj} 710, 903--923.
\newblock \doi{10.1088/0004-637X/710/2/903}
\input{Moster+10}

\bibitem[{{Naab} et~al.(2009){Naab}, {Johansson}, and {Ostriker}}]{Naab+09}
{Naab}, T., {Johansson}, P.~H., and {Ostriker}, J.~P. (2009).
\newblock {Minor Mergers and the Size Evolution of Elliptical Galaxies}.
\newblock \emph{\apjl} 699, L178--L182.
\newblock \doi{10.1088/0004-637X/699/2/L178}
\input{Naab+09}

\bibitem[{{Napolitano} et~al.(2005){Napolitano}, {Capaccioli}, {Romanowsky},
  {Douglas}, {Merrifield}, {Kuijken} et~al.}]{Napolitano+05}
{Napolitano}, N.~R., {Capaccioli}, M., {Romanowsky}, A.~J.,
{Douglas}, N.~G.,
  {Merrifield}, M.~R., {Kuijken}, K., et~al. (2005).
\newblock {Mass-to-light ratio gradients in early-type galaxy haloes}.
\newblock \emph{\mnras} 357, 691--706.
\newblock \doi{10.1111/j.1365-2966.2005.08683.x}
\input{Napolitano+05}

\bibitem[{{Napolitano} et~al.(2012){Napolitano}, {Capozziello}, {Romanowsky},
  {Capaccioli}, and {Tortora}}]{Napolitano+12_fR}
{Napolitano}, N.~R., {Capozziello}, S., {Romanowsky}, A.~J.,
{Capaccioli}, M.,
  and {Tortora}, C. (2012).
\newblock {Testing Yukawa-like Potentials from f(R)-gravity in Elliptical
  Galaxies}.
\newblock \emph{\apj} 748, 87.
\newblock \doi{10.1088/0004-637X/748/2/87}
\input{Napolitano+12_fR}

\bibitem[{{Napolitano} et~al.(2020){Napolitano}, {D'Ago}, {Tortora}, {Zhao},
  {Luo}, {Tang} et~al.}]{Napolitano+20_LAMOST}
{Napolitano}, N.~R., {D'Ago}, G., {Tortora}, C., {Zhao}, G.,
{Luo}, A.~L.,
  {Tang}, B., et~al. (2020).
\newblock {Central velocity dispersion catalogue of LAMOST-DR7 galaxies}.
\newblock \emph{\mnras} 498, 5704--5719.
\newblock \doi{10.1093/mnras/staa2409}
\input{Napolitano+20_LAMOST}

\bibitem[{{Napolitano} et~al.(2011){Napolitano}, {Romanowsky}, {Capaccioli},
  {Douglas}, {Arnaboldi}, {Coccato} et~al.}]{Napolitano+11_PNS}
{Napolitano}, N.~R., {Romanowsky}, A.~J., {Capaccioli}, M.,
{Douglas}, N.~G.,
  {Arnaboldi}, M., {Coccato}, L., et~al. (2011).
\newblock {The PN.S Elliptical Galaxy Survey: a standard {$\Lambda$}CDM halo
  around NGC 4374?}
\newblock \emph{\mnras} 411, 2035--2053.
\newblock \doi{10.1111/j.1365-2966.2010.17833.x}
\input{Napolitano+11_PNS}

\bibitem[{{Napolitano} et~al.(2010){Napolitano}, {Romanowsky}, and
  {Tortora}}]{NRT10}
{Napolitano}, N.~R., {Romanowsky}, A.~J., and {Tortora}, C.
(2010).
\newblock {The central dark matter content of early-type galaxies: scaling
  relations and connections with star formation histories}.
\newblock \emph{\mnras} 405, 2351--2371.
\newblock \doi{10.1111/j.1365-2966.2010.16710.x}
\input{NRT10}

\bibitem[{{Navarro} et~al.(1996){Navarro}, {Frenk}, and {White}}]{NFW96}
{Navarro}, J.~F., {Frenk}, C.~S., and {White}, S.~D.~M. (1996).
\newblock {The Structure of Cold Dark Matter Halos}.
\newblock \emph{\apj} 462, 563.
\newblock \doi{10.1086/177173}
\input{NFW96}

\bibitem[{{Nigoche-Netro} et~al.(2019){Nigoche-Netro}, {Ramos-Larios}, {Lagos},
  {de la Fuente}, {Ruelas-Mayorga}, {Mendez-Abreu} et~al.}]{Nigoche-Netro+19}
{Nigoche-Netro}, A., {Ramos-Larios}, G., {Lagos}, P., {de la
Fuente}, E.,
  {Ruelas-Mayorga}, A., {Mendez-Abreu}, J., et~al. (2019).
\newblock {The quantity of dark matter in early-type galaxies and its relation
  to the environment}.
\newblock \emph{\mnras} 488, 1320--1331.
\newblock \doi{10.1093/mnras/stz1786}
\input{Nigoche-Netro+19}

\bibitem[{{Nigoche-Netro} et~al.(2016){Nigoche-Netro}, {Ramos-Larios}, {Lagos},
  {Ruelas-Mayorga}, {de la Fuente}, {Kemp} et~al.}]{Nigoche-Netro+16}
{Nigoche-Netro}, A., {Ramos-Larios}, G., {Lagos}, P.,
{Ruelas-Mayorga}, A., {de
  la Fuente}, E., {Kemp}, S.~N., et~al. (2016).
\newblock {Dark matter inside early-type galaxies as function of mass and
  redshift}.
\newblock \emph{\mnras} 462, 951--959.
\newblock \doi{10.1093/mnras/stw1661}
\input{Nigoche-Netro+16}

\bibitem[{{Padmanabhan} et~al.(2004){Padmanabhan}, {Seljak}, {Strauss},
  {Blanton}, {Kauffmann}, {Schlegel} et~al.}]{Padmanabhan+04}
{Padmanabhan}, N., {Seljak}, U., {Strauss}, M.~A., {Blanton},
M.~R.,
  {Kauffmann}, G., {Schlegel}, D.~J., et~al. (2004).
\newblock {Stellar and dynamical masses of ellipticals in the Sloan Digital Sky
  Survey}.
\newblock \emph{\na} 9, 329--342.
\newblock \doi{10.1016/j.newast.2003.12.004}
\input{Padmanabhan+04}

\bibitem[{{Petrillo} et~al.(2019){Petrillo}, {Tortora}, {Vernardos},
  {Koopmans}, {Verdoes Kleijn}, {Bilicki} et~al.}]{Petrillo+19_LinKS}
{Petrillo}, C.~E., {Tortora}, C., {Vernardos}, G., {Koopmans},
L.~V.~E.,
  {Verdoes Kleijn}, G., {Bilicki}, M., et~al. (2019).
\newblock {LinKS: discovering galaxy-scale strong lenses in the Kilo-Degree
  Survey using convolutional neural networks}.
\newblock \emph{\mnras} 484, 3879--3896.
\newblock \doi{10.1093/mnras/stz189}
\input{Petrillo+19_LinKS}

\bibitem[{{Posti} et~al.(2014){Posti}, {Nipoti}, {Stiavelli}, and
  {Ciotti}}]{Posti+14}
{Posti}, L., {Nipoti}, C., {Stiavelli}, M., and {Ciotti}, L.
(2014).
\newblock {The imprint of dark matter haloes on the size and velocity
  dispersion evolution of early-type galaxies}.
\newblock \emph{\mnras} 440, 610--623.
\newblock \doi{10.1093/mnras/stu301}
\input{Posti+14}

\bibitem[{{Remus} et~al.(2013){Remus}, {Burkert}, {Dolag}, {Johansson}, {Naab},
  {Oser} et~al.}]{Remus+13}
{Remus}, R.-S., {Burkert}, A., {Dolag}, K., {Johansson}, P.~H.,
{Naab}, T.,
  {Oser}, L., et~al. (2013).
\newblock {The Dark Halo-Spheroid Conspiracy and the Origin of Elliptical
  Galaxies}.
\newblock \emph{\apj} 766, 71.
\newblock \doi{10.1088/0004-637X/766/2/71}
\input{Remus+13}

\bibitem[{{Renzini}(2006)}]{Renzini06}
{Renzini}, A. (2006).
\newblock {Stellar Population Diagnostics of Elliptical Galaxy Formation}.
\newblock \emph{\araa} 44, 141--192.
\newblock \doi{10.1146/annurev.astro.44.051905.092450}
\input{Renzini06}

\bibitem[{{Roy} et~al.(2018){Roy}, {Napolitano}, {La Barbera}, {Tortora},
  {Getman}, {Radovich} et~al.}]{Roy+18}
{Roy}, N., {Napolitano}, N.~R., {La Barbera}, F., {Tortora}, C.,
{Getman}, F.,
  {Radovich}, M., et~al. (2018).
\newblock {Evolution of galaxy size-stellar mass relation from the Kilo-Degree
  Survey}.
\newblock \emph{\mnras} 480, 1057--1080.
\newblock \doi{10.1093/mnras/sty1917}
\input{Roy+18}

\bibitem[{{Ruszkowski} and {Springel}(2009)}]{RS09}
{Ruszkowski}, M. and {Springel}, V. (2009).
\newblock {The Role of Dry Mergers for the Formation and Evolution of Brightest
  Cluster Galaxies}.
\newblock \emph{\apj} 696, 1094--1102.
\newblock \doi{10.1088/0004-637X/696/2/1094}
\input{RS09}

\bibitem[{{Saglia} et~al.(2010){Saglia}, {S{\'a}nchez-Bl{\'a}zquez}, {Bender},
  {Simard}, {Desai}, {Arag{\'o}n-Salamanca} et~al.}]{Saglia+10}
{Saglia}, R.~P., {S{\'a}nchez-Bl{\'a}zquez}, P., {Bender}, R.,
{Simard}, L.,
  {Desai}, V., {Arag{\'o}n-Salamanca}, A., et~al. (2010).
\newblock {The fundamental plane of EDisCS galaxies. The effect of size
  evolution}.
\newblock \emph{\aap} 524, A6.
\newblock \doi{10.1051/0004-6361/201014703}
\input{Saglia+10}

\bibitem[{{Salpeter}(1955)}]{Salpeter55}
{Salpeter}, E.~E. (1955).
\newblock {The Luminosity Function and Stellar Evolution.}
\newblock \emph{\apj} 121, 161.
\newblock \doi{10.1086/145971}
\input{Salpeter55}

\bibitem[{{Salucci}(2019)}]{Salucci+19_DM}
{Salucci}, P. (2019).
\newblock {The distribution of dark matter in galaxies}.
\newblock \emph{\aapr} 27, 2.
\newblock \doi{10.1007/s00159-018-0113-1}
\input{Salucci+19_DM}

\bibitem[{{S{\'a}nchez} et~al.(2012){S{\'a}nchez}, {Kennicutt}, {Gil de Paz},
  {van de Ven}, {V{\'{\i}}lchez}, {Wisotzki} et~al.}]{Sanchez+12_CALIFA_I}
{S{\'a}nchez}, S.~F., {Kennicutt}, R.~C., {Gil de Paz}, A., {van
de Ven}, G.,
  {V{\'{\i}}lchez}, J.~M., {Wisotzki}, L., et~al. (2012).
\newblock {CALIFA, the Calar Alto Legacy Integral Field Area survey. I. Survey
  presentation}.
\newblock \emph{\aap} 538, A8.
\newblock \doi{10.1051/0004-6361/201117353}
\input{Sanchez+12_CALIFA_I}

\bibitem[{{Sersic}(1968)}]{Sersic68}
{Sersic}, J.~L. (1968).
\newblock \emph{{Atlas de galaxias australes}} ("")
\input{Sersic68}

\bibitem[{{Shajib} et~al.(2020){Shajib}, {Treu}, {Birrer}, and
  {Sonnenfeld}}]{Shajib+21}
{Shajib}, A.~J., {Treu}, T., {Birrer}, S., and {Sonnenfeld}, A.
(2020).
\newblock {Massive elliptical galaxies at $z \sim 0.2$ are well described by
  stars and a Navarro-Frenk-White dark matter halo}.
\newblock \emph{arXiv e-prints} , arXiv:2008.11724
\input{Shajib+21}

\bibitem[{{Shu} et~al.(2015){Shu}, {Bolton}, {Brownstein}, {Montero-Dorta},
  {Koopmans}, {Treu} et~al.}]{Shu+15_SLACSXII}
{Shu}, Y., {Bolton}, A.~S., {Brownstein}, J.~R., {Montero-Dorta},
A.~D.,
  {Koopmans}, L.~V.~E., {Treu}, T., et~al. (2015).
\newblock {The Sloan Lens ACS Survey. XII. Extending Strong Lensing to Lower
  Masses}.
\newblock \emph{\apj} 803, 71.
\newblock \doi{10.1088/0004-637X/803/2/71}
\input{Shu+15_SLACSXII}

\bibitem[{{Simon} and {Geha}(2007)}]{Simon+07}
{Simon}, J.~D. and {Geha}, M. (2007).
\newblock {The Kinematics of the Ultra-faint Milky Way Satellites: Solving the
  Missing Satellite Problem}.
\newblock \emph{\apj} 670, 313--331.
\newblock \doi{10.1086/521816}
\input{Simon+07}

\bibitem[{{Smith} et~al.(2015){Smith}, {Lucey}, and
  {Conroy}}]{Smith+15_SINFONI}
{Smith}, R.~J., {Lucey}, J.~R., and {Conroy}, C. (2015).
\newblock {The SINFONI Nearby Elliptical Lens Locator Survey: discovery of two
  new low-redshift strong lenses and implications for the initial mass function
  in giant early-type galaxies}.
\newblock \emph{\mnras} 449, 3441--3457.
\newblock \doi{10.1093/mnras/stv518}
\input{Smith+15_SINFONI}

\bibitem[{{Sonnenfeld} et~al.(2013){Sonnenfeld}, {Treu}, {Gavazzi}, {Suyu},
  {Marshall}, {Auger} et~al.}]{Sonnenfeld+13_SL2S_IV}
{Sonnenfeld}, A., {Treu}, T., {Gavazzi}, R., {Suyu}, S.~H.,
{Marshall}, P.~J.,
  {Auger}, M.~W., et~al. (2013).
\newblock {The SL2S Galaxy-scale Lens Sample. IV. The Dependence of the Total
  Mass Density Profile of Early-type Galaxies on Redshift, Stellar Mass, and
  Size}.
\newblock \emph{\apj} 777, 98.
\newblock \doi{10.1088/0004-637X/777/2/98}
\input{Sonnenfeld+13_SL2S_IV}

\bibitem[{{Spiniello} et~al.(2012){Spiniello}, {Trager}, {Koopmans}, and
  {Chen}}]{Spiniello+12}
{Spiniello}, C., {Trager}, S.~C., {Koopmans}, L.~V.~E., and
{Chen}, Y.~P.
  (2012).
\newblock {Evidence for a Mild Steepening and Bottom-heavy Initial Mass
  Function in Massive Galaxies from Sodium and Titanium-oxide Indicators}.
\newblock \emph{\apjl} 753, L32.
\newblock \doi{10.1088/2041-8205/753/2/L32}
\input{Spiniello+12}

\bibitem[{{Swindle} et~al.(2011){Swindle}, {Gal}, {La Barbera}, and {de
  Carvalho}}]{SPIDER-V}
{Swindle}, R., {Gal}, R.~R., {La Barbera}, F., and {de Carvalho},
R.~R. (2011).
\newblock {SPIDER. V. Measuring Systematic Effects in Early-type Galaxy Stellar
  Masses from Photometric Spectral Energy Distribution Fitting}.
\newblock \emph{\aj} 142, 118.
\newblock \doi{10.1088/0004-6256/142/4/118}
\input{SPIDER-V}

\bibitem[{{Thomas} et~al.(2013){Thomas}, {Steele}, {Maraston}, {Johansson},
  {Beifiori}, {Pforr} et~al.}]{Thomas+13_BOSS}
{Thomas}, D., {Steele}, O., {Maraston}, C., {Johansson}, J.,
{Beifiori}, A.,
  {Pforr}, J., et~al. (2013).
\newblock {Stellar velocity dispersions and emission line properties of
  SDSS-III/BOSS galaxies}.
\newblock \emph{\mnras} 431, 1383--1397.
\newblock \doi{10.1093/mnras/stt261}
\input{Thomas+13_BOSS}

\bibitem[{{Thomas} et~al.(2009){Thomas}, {Saglia}, {Bender}, {Thomas},
  {Gebhardt}, {Magorrian} et~al.}]{ThomasJ+09}
{Thomas}, J., {Saglia}, R.~P., {Bender}, R., {Thomas}, D.,
{Gebhardt}, K.,
  {Magorrian}, J., et~al. (2009).
\newblock {Dark Matter Scaling Relations and the Assembly Epoch of Coma
  Early-Type Galaxies}.
\newblock \emph{\apj} 691, 770--782.
\newblock \doi{10.1088/0004-637X/691/1/770}
\input{ThomasJ+09}

\bibitem[{{Thomas} et~al.(2011){Thomas}, {Saglia}, {Bender}, {Thomas},
  {Gebhardt}, {Magorrian} et~al.}]{ThomasJ+11}
{Thomas}, J., {Saglia}, R.~P., {Bender}, R., {Thomas}, D.,
{Gebhardt}, K.,
  {Magorrian}, J., et~al. (2011).
\newblock {Dynamical masses of early-type galaxies: a comparison to lensing
  results and implications for the stellar initial mass function and the
  distribution of dark matter}.
\newblock \emph{\mnras} 415, 545--562.
\newblock \doi{10.1111/j.1365-2966.2011.18725.x}
\input{ThomasJ+11}

\bibitem[{{Tortora} et~al.(2007){Tortora}, {Cardone}, and
  {Piedipalumbo}}]{Tortora+07}
{Tortora}, C., {Cardone}, V.~F., and {Piedipalumbo}, E. (2007).
\newblock {Dynamical and gravitational lensing properties of a new
  phenomenological model of elliptical galaxies}.
\newblock \emph{\aap} 463, 105--118.
\newblock \doi{10.1051/0004-6361:20065552}
\input{Tortora+07}

\bibitem[{{Tortora} et~al.(2018{\natexlab{a}}){Tortora}, {Koopmans},
  {Napolitano}, and {Valentijn}}]{Tortora+18_Verlinde}
{Tortora}, C., {Koopmans}, L.~V.~E., {Napolitano}, N.~R., and
{Valentijn},
  E.~A. (2018{\natexlab{a}}).
\newblock {Testing Verlinde's emergent gravity in early-type galaxies}.
\newblock \emph{\mnras} 473, 2324--2334.
\newblock \doi{10.1093/mnras/stx2432}
\input{Tortora+18_Verlinde}

\bibitem[{{Tortora} et~al.(2012){Tortora}, {La Barbera}, {Napolitano}, {de
  Carvalho}, and {Romanowsky}}]{SPIDER-VI}
{Tortora}, C., {La Barbera}, F., {Napolitano}, N.~R., {de
Carvalho}, R.~R., and
  {Romanowsky}, A.~J. (2012).
\newblock {SPIDER - VI. The central dark matter content of luminous early-type
  galaxies: Benchmark correlations with mass, structural parameters and
  environment}.
\newblock \emph{\mnras} 425, 577--594.
\newblock \doi{10.1111/j.1365-2966.2012.21506.x}
\input{SPIDER-VI}

\bibitem[{{Tortora} et~al.(2014{\natexlab{a}}){Tortora}, {La Barbera},
  {Napolitano}, {Romanowsky}, {Ferreras}, and {de
  Carvalho}}]{Tortora+14_DMslope}
{Tortora}, C., {La Barbera}, F., {Napolitano}, N.~R.,
{Romanowsky}, A.~J.,
  {Ferreras}, I., and {de Carvalho}, R.~R. (2014{\natexlab{a}}).
\newblock {Systematic variations of central mass density slopes in early-type
  galaxies}.
\newblock \emph{\mnras} 445, 115--127.
\newblock \doi{10.1093/mnras/stu1616}
\input{Tortora+14_DMslope}

\bibitem[{{Tortora} et~al.(2016){Tortora}, {La Barbera}, {Napolitano}, {Roy},
  {Radovich}, {Cavuoti} et~al.}]{Tortora+16_compacts_KiDS}
{Tortora}, C., {La Barbera}, F., {Napolitano}, N.~R., {Roy}, N.,
{Radovich},
  M., {Cavuoti}, S., et~al. (2016).
\newblock {Towards a census of supercompact massive galaxies in the Kilo Degree
  Survey}.
\newblock \emph{\mnras} 457, 2845--2854.
\newblock \doi{10.1093/mnras/stw184}
\input{Tortora+16_compacts_KiDS}

\bibitem[{{Tortora} et~al.(2010{\natexlab{a}}){Tortora}, {Napolitano},
  {Cardone}, {Capaccioli}, {Jetzer}, and {Molinaro}}]{Tortora+10CG}
{Tortora}, C., {Napolitano}, N.~R., {Cardone}, V.~F.,
{Capaccioli}, M.,
  {Jetzer}, P., and {Molinaro}, R. (2010{\natexlab{a}}).
\newblock {Colour and stellar population gradients in galaxies: correlation
  with mass}.
\newblock \emph{\mnras} 407, 144--162.
\newblock \doi{10.1111/j.1365-2966.2010.16938.x}
\input{Tortora+10CG}

\bibitem[{{Tortora} et~al.(2009){Tortora}, {Napolitano}, {Romanowsky},
  {Capaccioli}, and {Covone}}]{Tortora+09}
{Tortora}, C., {Napolitano}, N.~R., {Romanowsky}, A.~J.,
{Capaccioli}, M., and
  {Covone}, G. (2009).
\newblock {Central mass-to-light ratios and dark matter fractions in early-type
  galaxies}.
\newblock \emph{\mnras} 396, 1132--1150.
\newblock \doi{10.1111/j.1365-2966.2009.14789.x}
\input{Tortora+09}

\bibitem[{{Tortora} et~al.(2010{\natexlab{b}}){Tortora}, {Napolitano},
  {Romanowsky}, and {Jetzer}}]{Tortora+10lensing}
{Tortora}, C., {Napolitano}, N.~R., {Romanowsky}, A.~J., and
{Jetzer}, P.
  (2010{\natexlab{b}}).
\newblock {Central Dark Matter Trends in Early-type Galaxies from Strong
  Lensing, Dynamics, and Stellar Populations}.
\newblock \emph{\apjl} 721, L1--L5.
\newblock \doi{10.1088/2041-8205/721/1/L1}
\input{Tortora+10lensing}

\bibitem[{{Tortora} et~al.(2018{\natexlab{b}}){Tortora}, {Napolitano}, {Roy},
  {Radovich}, {Getman}, {Koopmans} et~al.}]{Tortora+18_KiDS_DMevol}
{Tortora}, C., {Napolitano}, N.~R., {Roy}, N., {Radovich}, M.,
{Getman}, F.,
  {Koopmans}, L.~V.~E., et~al. (2018{\natexlab{b}}).
\newblock {The last 6 Gyr of dark matter assembly in massive galaxies from the
  Kilo Degree Survey}.
\newblock \emph{\mnras} 473, 969--983.
\newblock \doi{10.1093/mnras/stx2390}
\input{Tortora+18_KiDS_DMevol}

\bibitem[{{Tortora} et~al.(2014{\natexlab{b}}){Tortora}, {Napolitano},
  {Saglia}, {Romanowsky}, {Covone}, and {Capaccioli}}]{Tortora+14_DMevol}
{Tortora}, C., {Napolitano}, N.~R., {Saglia}, R.~P., {Romanowsky},
A.~J.,
  {Covone}, G., and {Capaccioli}, M. (2014{\natexlab{b}}).
\newblock {Evolution of central dark matter of early-type galaxies up to $z \sim 0.8$}.
\newblock \emph{\mnras} 445, 162--174.
\newblock \doi{10.1093/mnras/stu1712}
\input{Tortora+14_DMevol}

\bibitem[{{Tortora} et~al.(2018{\natexlab{c}}){Tortora}, {Napolitano},
  {Spavone}, {La Barbera}, {D'Ago}, {Spiniello} et~al.}]{Tortora+18_UCMGs}
{Tortora}, C., {Napolitano}, N.~R., {Spavone}, M., {La Barbera},
F., {D'Ago},
  G., {Spiniello}, C., et~al. (2018{\natexlab{c}}).
\newblock {The first sample of spectroscopically confirmed ultra-compact
  massive galaxies in the Kilo Degree Survey}.
\newblock \emph{\mnras} 481, 4728--4752.
\newblock \doi{10.1093/mnras/sty2564}
\input{Tortora+18_UCMGs}

\bibitem[{{Tortora} et~al.(2019){Tortora}, {Posti}, {Koopmans}, and
  {Napolitano}}]{Tortora+19_LTGs_DM_and_slopes}
{Tortora}, C., {Posti}, L., {Koopmans}, L.~V.~E., and
{Napolitano}, N.~R.
  (2019).
\newblock {The dichotomy of dark matter fraction and total mass density slope
  of galaxies over five dex in mass}.
\newblock \emph{\mnras} 489, 5483--5493.
\newblock \doi{10.1093/mnras/stz2320}
\input{Tortora+19_LTGs_DM_and_slopes}

\bibitem[{{Tortora} et~al.(2014{\natexlab{c}}){Tortora}, {Romanowsky},
  {Cardone}, {Napolitano}, and {Jetzer}}]{Tortora+14_MOND}
{Tortora}, C., {Romanowsky}, A.~J., {Cardone}, V.~F.,
{Napolitano}, N.~R., and
  {Jetzer}, P. (2014{\natexlab{c}}).
\newblock {MOND and IMF variations in early-type galaxies from ATLAS$^{3D}$}.
\newblock \emph{\mnras} 438, L46--L50.
\newblock \doi{10.1093/mnrasl/slt155}
\input{Tortora+14_MOND}

\bibitem[{{Tortora} et~al.(2013){Tortora}, {Romanowsky}, and
  {Napolitano}}]{TRN13_SPIDER_IMF}
{Tortora}, C., {Romanowsky}, A.~J., and {Napolitano}, N.~R.
(2013).
\newblock {An Inventory of the Stellar Initial Mass Function in Early-type
  Galaxies}.
\newblock \emph{\apj} 765, 8.
\newblock \doi{10.1088/0004-637X/765/1/8}
\input{TRN13_SPIDER_IMF}

\bibitem[{{Treu} et~al.(2010){Treu}, {Auger}, {Koopmans}, {Gavazzi},
  {Marshall}, and {Bolton}}]{Treu+10}
{Treu}, T., {Auger}, M.~W., {Koopmans}, L.~V.~E., {Gavazzi}, R.,
{Marshall},
  P.~J., and {Bolton}, A.~S. (2010).
\newblock {The Initial Mass Function of Early-Type Galaxies}.
\newblock \emph{\apj} 709, 1195--1202.
\newblock \doi{10.1088/0004-637X/709/2/1195}
\input{Treu+10}

\bibitem[{{Trujillo} et~al.(2004){Trujillo}, {Burkert}, and {Bell}}]{TBB04}
{Trujillo}, I., {Burkert}, A., and {Bell}, E.~F. (2004).
\newblock {The Tilt of the Fundamental Plane: Three-Quarters Structural
  Nonhomology, One-Quarter Stellar Population}.
\newblock \emph{\apjl} 600, L39--L42.
\newblock \doi{10.1086/381528}
\input{TBB04}

\bibitem[{{Trujillo} et~al.(2007){Trujillo}, {Conselice}, {Bundy}, {Cooper},
  {Eisenhardt}, and {Ellis}}]{Trujillo+07}
{Trujillo}, I., {Conselice}, C.~J., {Bundy}, K., {Cooper}, M.~C.,
{Eisenhardt},
  P., and {Ellis}, R.~S. (2007).
\newblock {Strong size evolution of the most massive galaxies since z \~{} 2}.
\newblock \emph{\mnras} 382, 109--120.
\newblock \doi{10.1111/j.1365-2966.2007.12388.x}
\input{Trujillo+07}

\bibitem[{{Trujillo} et~al.(2011){Trujillo}, {Ferreras}, and {de La
  Rosa}}]{Trujillo+11}
{Trujillo}, I., {Ferreras}, I., and {de La Rosa}, I.~G. (2011).
\newblock {Dissecting the size evolution of elliptical galaxies since $z \sim 1$: puffing-up versus minor-merging scenarios}.
\newblock \emph{\mnras} 415, 3903--3913.
\newblock \doi{10.1111/j.1365-2966.2011.19017.x}
\input{Trujillo+11}

\bibitem[{{Trujillo} et~al.(2006){Trujillo}, {F{\"o}rster Schreiber},
  {Rudnick}, {Barden}, {Franx}, {Rix} et~al.}]{Trujillo+06}
{Trujillo}, I., {F{\"o}rster Schreiber}, N.~M., {Rudnick}, G.,
{Barden}, M.,
  {Franx}, M., {Rix}, H.-W., et~al. (2006).
\newblock {The Size Evolution of Galaxies since z\~{}3: Combining SDSS, GEMS,
  and FIRES}.
\newblock \emph{\apj} 650, 18--41.
\newblock \doi{10.1086/506464}
\input{Trujillo+06}

\bibitem[{{Tulin} and {Yu}(2018)}]{Tulin_Yu18_SIDM}
{Tulin}, S. and {Yu}, H.-B. (2018).
\newblock {Dark matter self-interactions and small scale structure}.
\newblock \emph{\physrep} 730, 1--57.
\newblock \doi{10.1016/j.physrep.2017.11.004}
\input{Tulin_Yu18_SIDM}

\bibitem[{{van den Bosch} et~al.(2007){van den Bosch}, {Yang}, {Mo},
  {Weinmann}, {Macci{\`o}}, {More} et~al.}]{vdB+07}
{van den Bosch}, F.~C., {Yang}, X., {Mo}, H.~J., {Weinmann},
S.~M.,
  {Macci{\`o}}, A.~V., {More}, S., et~al. (2007).
\newblock {Towards a concordant model of halo occupation statistics}.
\newblock \emph{\mnras} 376, 841--860.
\newblock \doi{10.1111/j.1365-2966.2007.11493.x}
\input{vdB+07}

\bibitem[{{van der Wel} et~al.(2008){van der Wel}, {Holden}, {Zirm}, {Franx},
  {Rettura}, {Illingworth} et~al.}]{vanderWel+08}
{van der Wel}, A., {Holden}, B.~P., {Zirm}, A.~W., {Franx}, M.,
{Rettura}, A.,
  {Illingworth}, G.~D., et~al. (2008).
\newblock {Recent Structural Evolution of Early-Type Galaxies: Size Growth from
  z = 1 to z = 0}.
\newblock \emph{\apj} 688, 48--58.
\newblock \doi{10.1086/592267}
\input{vanderWel+08}

\bibitem[{Verlinde(2017)}]{Verlinde16}
Verlinde, E.~P. (2017).
\newblock {Emergent Gravity and the Dark Universe}.
\newblock \emph{SciPost Phys.} 2, 016.
\newblock \doi{10.21468/SciPostPhys.2.3.016}
\input{Verlinde16}

\bibitem[{{Viel} et~al.(2013){Viel}, {Becker}, {Bolton}, and
  {Haehnelt}}]{Viel+13}
{Viel}, M., {Becker}, G.~D., {Bolton}, J.~S., and {Haehnelt},
M.~G. (2013).
\newblock {Warm dark matter as a solution to the small scale crisis: New
  constraints from high redshift Lyman-{\ensuremath{\alpha}} forest data}.
\newblock \emph{\prd} 88, 043502.
\newblock \doi{10.1103/PhysRevD.88.043502}
\input{Viel+13}

\bibitem[{{Vogelsberger} et~al.(2020){Vogelsberger}, {Marinacci}, {Torrey}, and
  {Puchwein}}]{Vogelsberger+20_simulations}
{Vogelsberger}, M., {Marinacci}, F., {Torrey}, P., and {Puchwein},
E. (2020).
\newblock {Cosmological simulations of galaxy formation}.
\newblock \emph{Nature Reviews Physics} 2, 42--66.
\newblock \doi{10.1038/s42254-019-0127-2}
\input{Vogelsberger+20_simulations}

\bibitem[{{Wang} et~al.(2020){Wang}, {Vogelsberger}, {Xu}, {Mao}, {Springel},
  {Li} et~al.}]{WangY+20_IllustrisTNG}
{Wang}, Y., {Vogelsberger}, M., {Xu}, D., {Mao}, S., {Springel},
V., {Li}, H.,
  et~al. (2020).
\newblock {Early-type galaxy density profiles from IllustrisTNG - I. Galaxy
  correlations and the impact of baryons}.
\newblock \emph{\mnras} 491, 5188--5215.
\newblock \doi{10.1093/mnras/stz3348}
\input{WangY+20_IllustrisTNG}

\bibitem[{{Xu} et~al.(2017){Xu}, {Springel}, {Sluse}, {Schneider},
  {Sonnenfeld}, {Nelson} et~al.}]{Xu+17_Illustris}
{Xu}, D., {Springel}, V., {Sluse}, D., {Schneider}, P.,
{Sonnenfeld}, A.,
  {Nelson}, D., et~al. (2017).
\newblock {The inner structure of early-type galaxies in the Illustris
  simulation}.
\newblock \emph{\mnras} 469, 1824--1848.
\newblock \doi{10.1093/mnras/stx899}
\input{Xu+17_Illustris}

\bibitem[{{Zhan}(2018)}]{2018cosp...42E3821Z}
{Zhan}, H. (2018).
\newblock {An Overview of the Chinese Space Station Optical Survey}.
\newblock In \emph{42nd COSPAR Scientific Assembly}. vol.~42, E1.16--4--18
\input{2018cosp...42E3821Z}

\end{thebibliography}



\end{document}